\documentclass{aastex63}
\usepackage{bm}
\usepackage{ulem}
\usepackage{amsmath}
\usepackage{lineno}
\usepackage{color}
\usepackage{array}
\usepackage{tabu}
\usepackage{dcolumn}
\usepackage{natbib}

\usepackage{graphicx}
\graphicspath{{../pdf/}{../jpeg/}}

\begin{document}

\title{A fully self-consistent model for solar flares}

\author{Wenzhi Ruan}
\affiliation{Centre for mathematical Plasma Astrophysics, Department of Mathematics, KU Leuven, Celestijnenlaan 200B, B-3001 Leuven, Belgium}

\author{Chun Xia}
\affiliation{School of Physics and Astronomy, Yunnan University, Kunming 650050, China}

\author{Rony Keppens}
\affiliation{Centre for mathematical Plasma Astrophysics, Department of Mathematics, KU Leuven, Celestijnenlaan 200B, B-3001 Leuven, Belgium}

\received{Mar 27, 2020}
\revised{May 8, 2020}
\accepted{May 15, 2020}
\submitjournal{ApJ}

\begin{abstract}

The `standard solar flare model' collects all physical ingredients identified by multi-wavelength observations of our Sun: magnetic reconnection, fast particle acceleration and the resulting emission at various wavelengths, especially in soft to hard X-ray channels. Its cartoon representation is found throughout textbooks on solar and plasma astrophysics, and guides interpretations of unresolved energetic flaring events on other stars, accretion disks and jets. To date, a fully self-consistent model that reproduces the standard scenario in all its facets is lacking, since this requires the combination of a large scale, multi-dimensional magnetohydrodynamic (MHD) plasma description with a realistic fast electron treatment. Here, we demonstrate such a novel combination, where MHD combines with an analytic fast electron model, adjusted to handle time-evolving, reconnecting magnetic fields and particle trapping. This allows to study (1) the role of fast electron deposition in the triggering of chromospheric evaporation flows; (2) the physical mechanisms that generate various hard X-ray sources at chromospheric footpoints or looptops; and (3) the relationship between soft X-ray and hard X-ray fluxes throughout the entire flare loop evolution. For the first time, this self-consistent solar flare model demonstrates the observationally suggested relationship between flux swept out by the hard X-ray footpoint regions, and the actual reconnection rate at the X-point, which is a major unknown in flaring scenarios. We also demonstrate that a looptop hard X-ray source can result from fast electron trapping.

\end{abstract}


\section{introduction}
Solar flares are the most energetic explosions in our heliosphere, and power space weather events that affect our magnetosphere. Solar flares are known to efficiently convert magnetic energy into plasma internal and kinetic energy and emission. The `standard solar flare model' discussed in textbooks \citep{Tajima2002plap,Forbes2010hssr,Priest2014masu} extends the cartoon view of magnetic reconnection with many processes identified from observations: fast evaporation upflows, X-ray loops, and hard X-ray source regions. 
Hard X-ray (HXR) sources are frequently observed in flare events (e.g. \citealp{Masuda1994Natur,Su2013NatPh}). Observed HXR sources can be divided into looptop sources, appearing near the apex of the high temperature flare loop, and footpoint sources appearing near the chromospheric regions of the flare loop. The HXR sources, whether footpoint or looptop, are believed to be generated via bremsstrahlung mechanisms: fast electrons interact or collide with background charged particles and release X-ray photons (e.g. \citealp{Kontar2011SSRv}). These fast electrons are  produced in a reconnection site near the flare loop apex, via direct electric field acceleration, stochastic acceleration or shock acceleration \citep{Aschwanden2005psci}. Accelerated fast electrons interact with plasma near the looptop and can produce a looptop HXR source, and they generally move along the magnetic field to lose their energy in the denser chromosphere via collisions to produce footpoint HXR sources. The HXR flux is found to have a close relationship with the soft X-ray (SXR) flux, implying that the temporal profile of the HXR flux loosely fits the time derivative of the SXR profile. This observational finding is known as the Neupert effect \citep{Neupert1968ApJ,Hudson1991BAAS}. This Neupert effect suggests a close relationship between the thermal and non-thermal energy evolution: the HXR flux indicates the transfer rate from non-thermal energy to thermal energy due to collisions, while the SXR flux indicates the total thermal energy of the flare loop.
HXR emission and the related fast electrons in flare events have been studied for several decades, but they still pose many questions to be answered theoretically. For example: (1) how exactly are looptop HXR sources produced; or (2) how great is the contribution of the fast electron energy deposition to the triggering of chromospheric evaporation flows? Since both questions involve interaction between fast electrons and the background plasma, they are difficult to study with magnetohydrodynamic (MHD) simulations alone. A complete macroscopic model for the standard solar flare scenario is lacking, since this requires coupling of MHD with physics related to kinetic processes, like the generation of fast electron beams, and their subsequent interaction with the plasma and radiation fields.

For the HXR sources, it is difficult to imagine how sources which are comparable to the chromospheric footpoint sources, can be generated in the low density looptops, since the efficiency of bremsstrahlung is determined by the local plasma density. \citet{Krucker2008A&ARv} suggests that some observed flare loops behave as a thick target for fast electrons, such that few fast electrons arrive at the chromosphere, but this idea has not been confirmed by numerical simulations. A looptop HXR source has been reproduced in a recent three dimensional (3D) radiative MHD simulation that is triggered by flux emergence in \citet{Cheung2019NatAs}. However, the contribution of non-thermal electrons is not included and their strong looptop emission results from unrealistically hot plasma, reaching temperatures beyond 100 MK. Moreover, this scenario can not explain the generation of the HXR sources observed at chromospheric footpoints. For the question related to evaporation flows, numerical studies conducted by \citet{Fisher1985ApJ} show that fast electron beams can produce evaporation flows of several hundred $\rm km\ s^{-1}$  in seconds when the energy flux carried by the beams reaches $5\times 10^{10} \ \rm erg\ cm^{-2}\ s^{-1}$. However, 2D magnetohydrodynamic (MHD) simulations in \citet{Yokoyama2001ApJ} show that thermal conduction is also a possible way to produce chromospheric evaporation. It is therefore still unknown which mechanism determines or dominates the evaporation process, although restricted 1D hydro models do include and compare thermal conduction with fast electron treatments (e.g. pioneering work by \citealp{Abbett1999ApJ,Allred2005ApJ}). Since both questions from above involve interaction between fast electrons and the multi-dimensionally evolving background plasma, they are difficult to study with MHD simulations alone. Kinetic models employing particle-in-cell (PIC) viewpoints are suitable to study microscopic behaviors related to fast electron beams, but it is numerically challenging to reproduce the entire evolution of a macroscopic flare with these methods. A combination of MHD and PIC has been demonstrated by \citet{Baumann2013ApJ} to study the behaviour of fast electrons in a solar flare, in which the MHD simulation provides initial and boundary conditions for the PIC simulation. Their study focuses on the acceleration of the fast electrons in reconnection null-point regions and there is no feedback from the PIC simulation to the MHD evolution. In this paper, we pioneer a numerically ingenious way to study the interaction, by combining MHD simulations with analytic fast electron models.

Analytic fast electron models aim to estimate the change of the fast electron energy and pitch angle owing to collisions via analytic means, where it is assumed that these electrons are moving along magnetic field lines. Two types of models are often used in solar physics, one based on the solution of the Boltzmann equation with a Fokker-Planck treatment of collisions  \citep{McTiernan1990ApJ,Allred2015ApJ,Kerr2016ApJ,Brown2018ApJ,Emslie2018ApJ}, the other based on analytic energy loss rates of test electron beams in cold target plasma using a scattering treatment \citep{Brown1972SoPh,Emslie1978ApJ}. Analytic fast electron models, together with one dimensional (1D) (M)HD models, have been widely used to study the generation of evaporation of chromospheric plasma due to energetic electron deposition (e.g. \citealp{Fisher1985ApJ,Abbett1999ApJ,Allred2005ApJ}). Recently, \citet{Ruan2019ApJ} employed for the first time such analytic electron model in a two-and-a-half dimensional (2.5D) MHD simulation, in which the electron deposition processes along multiple, but fixed magnetic field lines were reproduced, and evaporation processes were demonstrated. These numerical studies clearly show that the penetration of fast electrons can indeed lead to evaporation of chromospheric plasma when the energy flux carried by the fast electrons is high enough. The fast electrons in these studies are assumed to have a given initial single or double power law spectrum and the temporal energy fluxes carried by the electrons down the field lines are then analyzed. Another approach is pursued in \citet{Bakke2018A&A}, linking energy flux to purely numerical reconnection to get a temporal and spatial energy flux redistribution in their 3D radiative MHD simulations. There, fast electron effects are included based on a Fokker-Planck equation, but their simulation time covers only 10 solar seconds, and has no clearly resolved reconnection layers, so this study did not answer the solar flare questions posed earlier, e.g. those concerning the role of evaporation flows. 

Here we combine a multi-dimensional MHD model with a dynamic, multi-dimensional analytic fast electron model to study HXR emission and chromospheric evaporation in solar flares. The magnetic reconnection process is resolved in the MHD model, and provides the energy to the fast electrons. The fast electrons are assumed to become accelerated in the reconnection current sheet and move downwards to the lower atmosphere along magnetic field lines. As in \citet{Bakke2018A&A}, the temporal and spatial flux of the fast electrons are linked to the reconnection process, and energy is fed back in a non-local, time-evolving fashion to the plasma described by the MHD equations. Especially novel in our approach is the possibility to address looptop particle trapping. The method and setup of our simulation is introduced in section~\ref{Setup}. The results are displayed and discussed in section~\ref{result}, in which the reasons for the chromospheric evaporations and the mechanism for the generation of looptop HXR sources are analyzed. Conclusions are provided in section~\ref{Concl}.

\section{The numerical setup}\label{Setup}

\subsection{Governing equations}

The simulation is performed with the open-source MPI-parallelized Adaptive Mesh Refinement Versatile Advection Code \citep{Keppens2012JCoPh, Porth2014ApJS, Xia2018ApJS}. The simulation is two-and-a-half dimensional (2.5D) where all vector quantities have three components, while showing variations on a two dimensional domain of $-50\ \textrm{Mm} \leq x \leq 50\ \textrm{Mm}$ and $0\ \textrm{Mm} \leq y \leq 100\ \textrm{Mm}$. The governing equations of the simulation are given by
\begin{eqnarray}
\frac{\partial \rho}{\partial t} + \nabla \cdot (\rho \bm{v}) & = & 0, \\ 
\frac{\partial \rho \bm{v}}{\partial t} + \nabla \cdot (\rho \bm{v} \bm{v} + p_{\rm{tot}} \bm{I} - \bm{BB}) & = & \rho \bm{g}, \\
\frac{\partial e}{\partial t} + \nabla \cdot (e \bm{v} + p_{\rm{tot}} \bm{v} - \bm{BB} \cdot \bm{v}) & = & \rho \bm{g} \cdot \bm{v} +
\nabla \cdot (\bm{\kappa} \cdot \nabla T) + \nabla \cdot (\bm{B} \times \eta \bm{J}) - Q_r - Q_e + H_b + H_e, \label{q-en} \\
\frac{\partial \bm{B}}{\partial t} + \nabla \cdot (\bm{v B} - \bm{B v}) & = & - \nabla \times (\eta \bm{J}) \,,
\end{eqnarray}
where $\rho$, $\bm{v}$, $p$, $\bm{B}$, $e$, $T$, $p_{tot}$ and $\bm{J}= \nabla \times \bm{B}$ are plasma density, velocity, pressure, magnetic field, total energy density, temperature, total pressure and current density respectively. Furthermore, $\bm{I}$ is the unit tensor, $\bm{\kappa}=  \kappa_{\parallel} T^{5/2} \bm{\hat{b}}\bm{\hat{b}}$ is the adopted thermal conductivity tensor where $\kappa_{\parallel} = 8 \times 10^{-7} \ \rm {erg\ cm^{-1}\ s^{-1}\ K^{-7/2}}$ and $\bm{\hat{b}}$ is a unit-tangent to the magnetic field vector, while $\eta(x,y,t)$ is the resistivity. The gravitational stratification enters as $\bm{g}=-274 R_{\rm{s}}^2 / (R_{\rm{s}} + y)^2 \bm{\hat{y}} \ \rm m\ s^{-2}$ where $R_{\rm{s}}$ denotes the solar radius. Note that we solve these equations in dimensionless units, where the permeability of free space no longer enters. An ideal gas law closes the system where the ratio of specific heats is $\gamma=5/3$.

The energy equation~(\ref{q-en}) contains at its far right four additional gain-loss terms, where two are specific to the coronal plasma conditions ($Q_r$, $H_b$), while two others relate to the fast electrons ($Q_e$ and $H_e$).
$Q_r=N_e N_p \Lambda(T)$ is the optically thin radiative cooling effect that is prominent in all coronal plasma, and is calculated with a cooling curve $\Lambda(T)$ provided by \cite{Colgan2008ApJ}. Electron and proton number densities in a fully ionized hydrogen plasma obey $N_e=N_p=\rho/m_p$ where $m_p$ indicates the proton mass, and the MHD approach assigns an equal temperature to both species $T=T_e=T_p$. The background heating $H_b$ is parametrically given by 
\begin{equation}
H_b = \max \left \{ \frac{ c_0 [(y-y_0)/h_0]^{-2/7}}{\exp[h_1/(y-y_1)] - 1} ,\ 0 \right \},
\end{equation}
where $c_0=0.01\ \rm erg\ cm^{-3}\ s^{-1}$, $y_0=1\ \rm Mm$, $h_0=5 \ \rm Mm$, $y_1=2 \ \rm Mm$, $h_1=3 \ \rm Mm$. This background heating, which has a peak value at our transition region roughly located at about 2.543 Mm, is used to compensate the strong radiative cooling in the transition region. 

The influence of non-thermal electrons has been included by means of the $Q_e$ and $H_e$ terms. $Q_e$ is the energy used to accelerate the electrons, and relates to the anomalous resistivity as described in Section~\ref{ares}, where we assume that the electrons are accelerated in the reconnection current sheet and locally consume a fraction of the Ohmic heating in their energization. They finally deposit and lose their energy due to collisions when moving along magnetic field lines. Thereby, $H_e$ becomes a non-local, time-dependent heating term owing to collisions between fast electrons and the background plasma. This term allows for handling the possibility of fast particle trapping, and quantifies energy deposition rates in time-varying, multi-dimensional magnetic field configurations. Its implementation details are provided in Appendix~\ref{app1}. 

An initial resolution of $64 \times 64$ and a maximum refinement level of 6 are adopted. The equivalent resolution of $2048\times 2048$ is achieved due to adaptive mesh refinement (AMR). We hence have a resolution down to 48 km per grid cell. The HLL approximate Riemann solver and a mixture of high order slope limiters (third order limiter presented by \citet{Cada2009JCoPh} and second order limiter presented by \citet{vanLeer1974JCoPh}) are employed in our simulation. The second order limiter is employed for the lower atmosphere, reconnection region and flare loop whenever the refinement level is higher than 3, while the third order limiter is employed in the background corona where the refinement level is less or equal to 3.

\subsection{Initial conditions and boundary conditions}\label{initbc}

The initial configuration of the atmospheric magnetic field is similar to that in \citet{Yokoyama2001ApJ}, which is essentially a vertical current sheet given by
\begin{eqnarray}
B_x &=& 0, \\
B_y &=& B_0 \tanh (x/w), \\
B_z &=& B_0 / \cosh (x/w),
\end{eqnarray}
where $B_0 = 35 \ \rm G$ and $w=1.5 \ \rm Mm$. The VAL-C temperature profile is employed as initial temperature profile in the region below the transition region height of $h_{\rm{tra}} =2.543 \ \rm Mm$ \citep{Vernazza1981ApJS}. The temperature profile above $h_{\rm{tra}}$ is given by
\begin{equation}
T(y) = [3.5 F_{\rm{c}} (y - h_{\rm{tra}}) / \kappa_{\parallel} + T_{\rm{tra}}^{7/2}]^{2/7},
\end{equation}
where $F_{\rm{c}} = 1.2 \times 10^6 \ \rm erg \ cm^{-2} \ s^{-1}$ and $T_{\rm{tra}} = 0.447 \ \rm MK$. The number density at the lower boundary is $4.6 \times 10^{14} \ \rm cm^{-3}$ and the initial density $\rho(y,t=0)$ is calculated based on hydrostatic equilibrium. The coronal number density has a value of $3.3 \times 10^9 \ \rm cm^{-3}$ at $y=10\ \rm Mm$ and a value of $0.9 \times 10^9 \ \rm cm^{-3}$ at the upper boundary $y=100\ \rm Mm$. The coronal plasma beta thereby varies from $0.028$ at $y=10\ \rm Mm$ to $0.016$ at $y=100\ \rm Mm$.

At the left and right boundaries, symmetric conditions are employed for $\rho$, $v_y$, $p$ and $B_y$, while $v_x$, $v_z$, $B_x$ and $B_z$ apply asymmetric conditions.   At the bottom boundary, $\rho$, $p$ and $\bm{B}$ are fixed and asymmetric conditions are employed for $\bm{v}$. At the upper boundaries, an asymmetric condition is used for $B_x$ to impose that the field-aligned heat flux is perpendicular to the boundary. The temperature at the upper boundary is set according to $dT/dy=0$ when the temperature $T_b<5\ \rm MK$, where $T_b$ is the instantaneous local temperature value at the upper boundary. As explained in Section~\ref{ares}, our anomalous resistivity varies as based on flow speeds and temperature, and in the reconnection upflow region we will find $T_b>5 \ \rm MK$ at the upper boundary. In such locations, we impose $dT/dy=-T_b/(20\ \rm Mm)$ to avoid runaway high temperatures at the upper, open boundary that could introduce prohibitively small simulation time steps. All other variables are free for this boundary, hence are extrapolated with Neumann conditions. As we especially are interested in the flare effects on chromospheric regions, we will always focus on the reconnection downflow region. Since this is far from the upper boundary, it is not influenced by our upper boundary treatment.

\subsection{Resistivity and fast electron penetration}\label{ares}

We use a two-stage, spatio-temporal anomalous resistivity model as in \citet{Yokoyama2001ApJ}. The initial reconnection process is triggered via providing a spatially localized resistivity
\begin{equation}
\eta (x,y,t<t_{\eta}) = 
\begin{cases}
\eta_0 [2 (r/r_{\eta})^3 - 3 (r/r_{\eta})^2 + 1] , & \quad r \leq r_{\eta} \\
0 , & \quad r>r_{\eta}
\end{cases}
\end{equation}
where $\eta_0=5\times 10^{-2}$, $r=\sqrt{x^2+(y-h_{\eta})^2}$, $h_{\eta}=50 \ \rm Mm$, $r_{\eta}=2.4 \ \rm Mm$ and $t_{\eta}=31.2 \ \rm s$.  This profile just serves to locate the initial X-point at the controlled height $h_{\eta}$. In a second stage, anomalous resistivity (physically caused by microscopic instabilities that are unresolved in MHD) is employed for $t\geq t_{\eta}$, which reads
\begin{equation}
\eta (x,y,t\geq t_{\eta}) = 
\begin{cases}
0 , & \quad v_d \leq v_c \\
\textrm{min}\{ \alpha_{\eta} (v_d/v_c-1) \exp[(y-h_{\eta})^2/h_s^2] , 0.1\}, & \quad v_d > v_c
\end{cases}
\end{equation}
where $\alpha_{\eta}=1\times 10^{-4}$ and $h_s=10 \ \rm Mm$. Here, the instantaneous distribution of $v_d(x,y,t)=J/(e N_e)$ quantifies the relative ion-electron drift velocity, where the electron charge $-e$ enters. This activates anomalous resistivity when the drift velocity exceeds the treshold $v_c=1000\ u_{v} $, where $u_{v}\approx 128 \ \rm km\ s^{-1}$ is the unit of velocity in our simulation used for non-dimensionalization.

Non-thermal electrons are included in our model to quantify chromospheric heating due to their energy deposition, but their evolution is not described by MHD equations. We handle the fast electrons based on instantaneous magnetic field line tracing and adopt a particle treatment that significantly modifies and generalizes the 1D treatment of \citet{Emslie1978ApJ} for fast electron beams through multi-dimensional, time-evolving field lines, along with allowing for particle trapping. The method to trace magnetic field lines through our AMR grid is introduced in Appendix~\ref{app2}. Multiple magnetic field lines are dynamically located through the grid by magnetic field line tracing, and we interpolate all relevant quantities from the grid to the field lines and back, as explained in Appendix~\ref{app4}. 
The local generation and evolution of the fast electrons through the field lines employs a physics-based model for fast electron beams interacting with background plasma, building in the first adiabatic invariancy associated with particle trapping, as explained in detail in Appendices~\ref{app1} and \ref{app3}. Fast electrons are assumed to be accelerated in the downward reconnection outflows only, as we focus on the chromospheric flare response. The rate of total energy transfer to the fast electrons is given by the local Joule heating, in particular
\begin{equation}
Q_e (x,y,t\geq t_{\eta}) = 
\begin{cases}
 \eta J^2,   & v_y < - u_{v}\ \& \  T > 5 \ \rm MK \\
 0,  &  \ \rm elsewhere.
\end{cases}
\label{joule}
\end{equation}
Every field line traced is actually treated as a dynamical flux tube element, allowing to generalize concepts like the continuity of the fast electron flux through the field line bundle to 2D or 3D situations. We dynamically integrate $Q_e$ through the tube element to quantify the instantaneously consumed total energy for newly accelerated electrons. We dynamically update the beam-averaged pitch angle distribution along the flux tube and compute the energy flux and budget for the moving fast electrons through the evolving and relocating tube element. The energy flux then employs a new 1D analytic fast electron model introduced in Appendix~\ref{app1} to calculate the heating rate due to collisions along the field line. Thereafter, interpolation is performed from fieldlines to grid, to obtain the heating rate to the finite volume grid cells employed in the MHD simulation (i.e., the source term $H_e$ in the governing equations of MHD). The interpolation method can be found in Appendix~\ref{app4}.

To employ the analytic fast electron model in dynamically evolving flux tubes, we introduce and track plasma and beam properties at a reference point along each field line, as explained in Appendix~\ref{app1}, so that we can always integrate plasma density along the magnetic field line and quantify column depths. The heating rate in the model is a function of column depth, and our reference point arbitrary sets the local column depth to zero. The expected symmetry of the evolution, in accord with the initial left-right symmetry, allows us to choose the point where $x=0$, located at the middle of a typical field line, as the reference point. We assume that fast electrons move along field lines from the reference point to both (left and right) chromospheric footpoints. The fast electrons have a dynamically evolving input energy flux $\mathcal{F}_0(t)$, cosine pitch angle $\mu_0(t)$, and a fixed power law spectrum of spectral index $\delta=4$ and cutoff energy $E_c=20\ \rm keV$ at the reference point. The cosine pitch angle distribution along each field line is calculated based on the assumption that fast electrons ejected from the acceleration site have a fixed average pitch angle where $\arcsin (v_{\perp,a}/v_{e,a}) =18^{\circ}$, where $v_{\perp,a}$ is the average perpendicular velocity of the fast electrons and $v_{e,a}=10^5 \ \rm km\ s^{-1}$ is the average speed of the newly accelerated electrons. The evolution of the pitch angle is governed by the first adiabatic invariant $v_{\perp}^2(s,t)/B(s,t)=const$ and the effect of particle trapping can be fully reproduced. The energy flux $\mathcal{F}_0(t)$ is obtained by accounting for the total energy budget evolution of the newly accelerated particles, incorporating their gains from and losses to the MHD field, while accounting for trapping. More details about the calculations involved to update $\mathcal{F}_0(t)$ and $\mu_0(t)$ are available in Appendix~\ref{app3}. Relativistic effects are not included in our analytic electron model, in accordance with the typical average fast electron beam speeds $v_{e,a}$ associated with the cut-off energy $E_c$, which have a Lorentz factor of 1.06.

\section{Results}\label{result}

\subsection{Evolution of the flare loop}

\begin{figure*}[htbp]
\begin{center}
\includegraphics[width=0.8\linewidth]{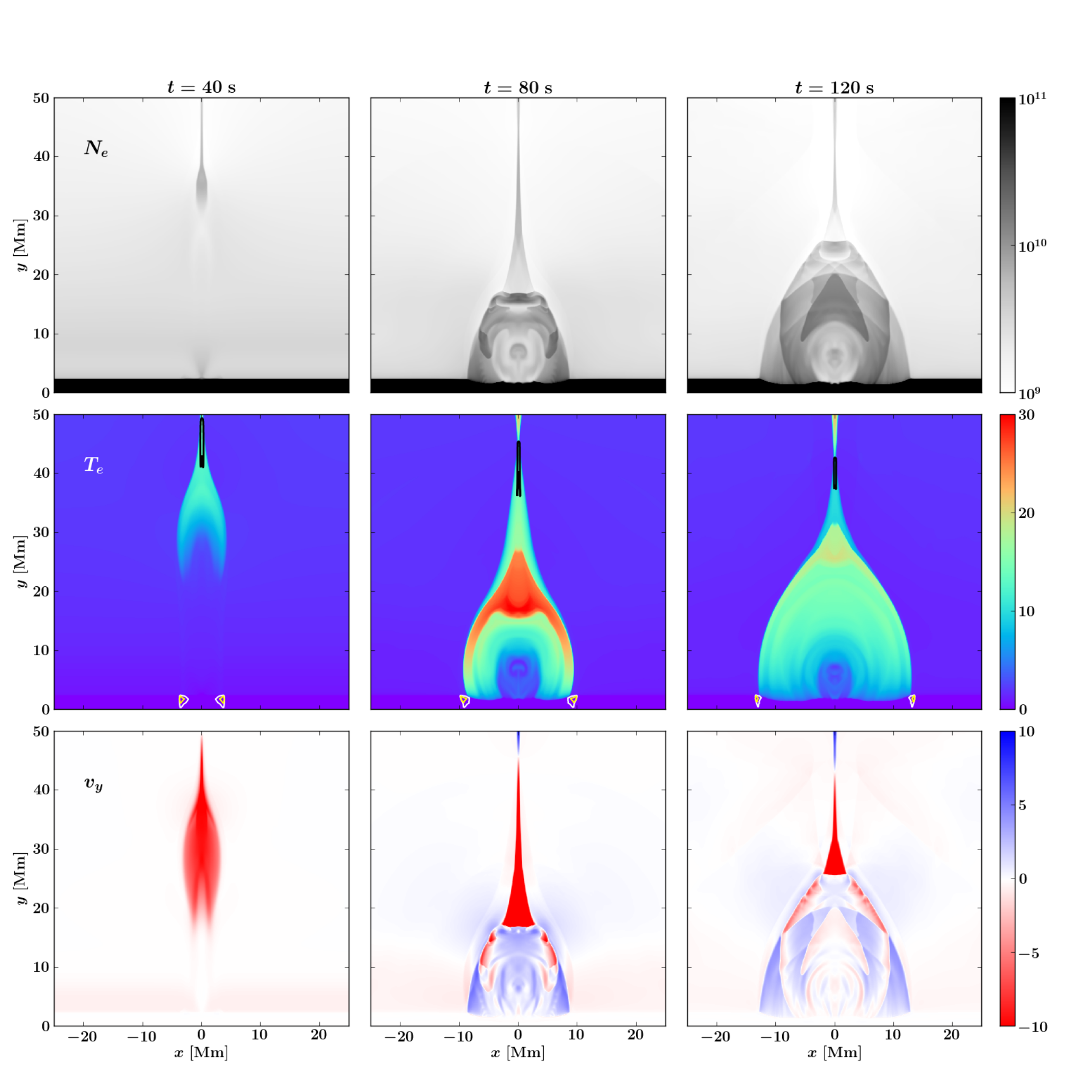}
\caption{Number density $N_e$ (in cm$^{-3}$), temperature $T_e$ (in MK) and vertical velocity $v_y$ (in 100 km s$^{-1}$) at $t=$40, 80 and 120 $\rm s$.  In the temperature views (middle row),  white and yellow contours near the flare loop footpoints show the heating due to fast electron energy deposition, with a level of $1\%$ and $10\%$ of the maximum values, respectively. In the same panels, the black contour identifies the instantaneous region of fast electron energization.}
\label{global}
\end{center}
\end{figure*}

Figure~\ref{global} shows the temporal evolution of number density, temperature and vertical velocity. The evolution is well described by the standard flare model: 1) a hot flare loop is formed owing to reconnection; 2) high speed reconnection outflows collide with the flare loop and produce a shock above the loop; 3) high density evaporation flows are produced and fill the flare loop. The speed of the outflow, up to $1600 \ \rm km\ s^{-1}$, leads to a high temperature of $30\ \rm MK$ downstream of the shock. The temperature in the rest of the loop is about $20\ \rm MK$, which is a temperature often reported in flare observations (e.g. \citealp{Masuda1994Natur}). The density of the evaporation flow is $\sim 10^{10}\ \rm cm^{-3}$ and the velocity of this flow is $\sim 500 \ \rm km\ s^{-1}$. The mechanism of the evaporation will be studied in the following section. The fast electron acceleration site and the associated energy deposition sites are marked by black and white-yellow contours, respectively, on the temperature panels. It clearly shows that the fast electrons get their energy from the outflow current sheet region and subsequently lose their energy in the chromosphere due to collisions. The energy deposition rate at $t=$40, 80 and 120 s, which is obtained via integrating the heating rate (erg cm$^{-3}$ s$^{-1}$) at the chromosphere in our 2D simulation box, is $3.8 \times 10^{17}$, $2.8 \times 10^{17}$ and $8.4 \times 10^{16} \ \rm erg\ cm^{-1}\ s^{-1}$ respectively. The total energy deposition rate then works out to be about $10^{28}\ \rm erg\ s^{-1}$ if we assume that the width of the flare loop in the (invariant) $z$-direction is 20-30 Mm, which is a general penetration rate that is frequently reported (e.g. \citealp{Veronig2005ApJ}). Note that the simulation box is bigger than the region shown in Figure~\ref{global}, since we only focus on the reconnection region and flare loop.

\subsection{Evaporation of plasma}

It has been proven by numerical simulations that both fast electron penetration and thermal conduction can lead to the evaporation of chromospheric plasma (e.g. \citealp{Fisher1985ApJ,Yokoyama2001ApJ}). Here we combine multi-dimensional anisotropic thermal conduction in time dependent MHD and fast electrons to study and compare the contributions of the two mechanisms to the evaporation.

\begin{figure}[htbp]
\begin{center}
\includegraphics[width=\linewidth]{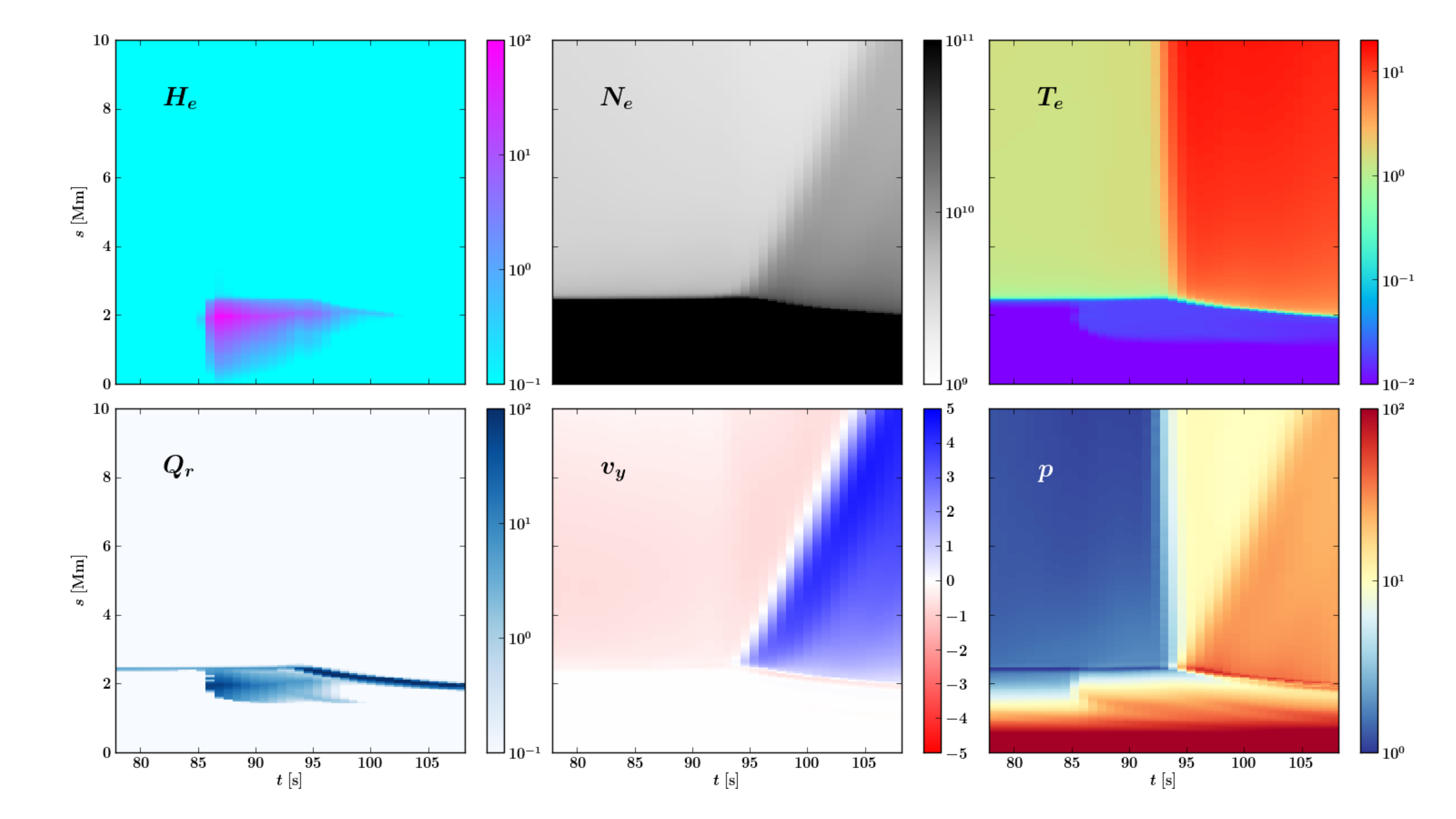}
\caption{Time-space plots of electron deposition heating rate $H_e(s,t)$ (erg cm$^{-3}$ s$^{-1}$), number density $N_e(s,t)$ (cm$^{-3}$), temperature $T_e(s,t)$ (MK), radiative cooling $Q_r(s,t)$ (erg cm$^{-3}$ s$^{-1}$), vertical velocity $v_y(s,t)$ (100 km s$^{-1}$), and thermal pressure $p(s,t)$ (dyne cm$^{-2}$) along a selected magnetic field line. The vertical $s$-axis indicates the distance along the field line.}
\label{field}
\end{center}
\end{figure}

Where the fast electrons lose their energy can be deduced from Figure~\ref{global}: their energy is deposited to those regions where the evaporation has not yet happened, spreading laterally away from the symmetry axis as flare ribbons. In between, we find the region where the evaporation is strong, which grows in size as the flare proceeds. The temporal evolution of plasma parameters along a pre-selected magnetic field line is also analyzed, to understand the relationship between fast electron penetration and chromospheric evaporation. The magnetic field line we selected is always traced from its footpoint, which is located at the lower boundary where $x=10$ Mm and $y=0$ Mm. As time progresses, this field line becomes part of the flare loop system. Time-space plots of the heating rate owing to electron penetration, the density, temperature and upward velocity are illustrated in Figure~\ref{field}. The energy flux into the chromosphere can be obtained via integrating the heating rate (erg cm$^{-3}$ s$^{-1}$) in the chromosphere along the field line. The fast electron beams arrive at the chromosphere at $\sim 85$ s,  reach an energy flux maximum of $\sim 3.4 \times 10^9 \ \rm erg\ cm^{-2}\ s^{-1} $ at $\sim 86 \ \rm s$ and then the energy flux decreases to e-fold from its peak value in 3 seconds. The heating due to fast electron deposition and the evaporation do not happen in phase: the heating appears earlier than the evaporation and the heating has essentially disappeared when the evaporation starts. Moreover, the chromospheric plasma evaporates continuously even while there is no more fast electron heating. In contrast, the evaporation does have a strong relationship with thermal conduction: the temperature shows an obvious gradient along the field line when the evaporation starts. Pressure has a local maximum at the upper transition region where the temperature is close to $1\ \rm MK$ ($y$ is about 2-3 Mm), when the downward heat flux has reached the lower atmosphere and the low corona reaches high temperatures of $\sim$10 MK. Obvious upward speeds can be found for the region above where the pressure has a local peak value. All of the evidence indicates that the evaporation of plasma is caused by thermal conduction rather than fast electron penetration in this simulation. Therefore, we suggest that thermal conduction is here more efficient than fast electron deposition in producing plasma evaporation. 
 
The main reason for the low efficiency of fast electron penetration in triggering evaporation is that fast electrons tend to lose energy at the chromosphere where the plasma is heated to about $2\times10^4$ K and density is still very high, leading to very strong optically thin radiative cooling which efficiently carries away the deposited energy from the fast electrons. About $70\%$ of the deposited energy is lost due to radiative cooling. The penetration leads to an increase of chromospheric temperature and pressure, but no obvious evaporation flow is produced. Our result supports the conclusion of \citet{Fisher1985ApJ} that very little chromospheric evaporation occurs at an energy flux of order $10^9 \ \rm erg\ cm^{-2}\ s^{-1}$. According to the survey in \citet{Fisher1985ApJ}, an energy flux higher than $10^{10} \ \rm erg\ cm^{-2}\ s^{-1}$ is required to drive obvious chromospheric evaporation.

The evaporation in our simulation tends to be transition region evaporation rather than chromospheric evaporation, since the upward flow is triggered from where the temperature is already in the upper transition region. This phenomenon is consistent with the Doppler shift analysis of EUV lines of an observed flare event in \citet{Li2011ApJ}, in which blue shift starts to appear at the emission lines where the corresponding temperature is close to $1\ \rm MK$. The reason for this evaporation is the accumulation of energy due to the rapid decrease of thermal conduction efficiency when the heat flux flows from the flare loop to the chromospheric footpoints, since this efficiency is determined by the local temperature. The previous balance between thermal conduction and radiative loss in the lower solar atmosphere has been broken, since the density and temperature has been modified by the strong heat flux. The phenomenon that blue shifts start to appear at a high temperature, much higher than the typical temperature of the chromosphere or low transition region, is also reported in \citet{Milligan2009ApJ}. Blue shifts were observed in the emission lines from Fe XIV to Fe XXIV (2–16 MK) and red shifts were found in the emission lines from He II to Fe VIII (0.05–1.5 MK) in their case study of a C-class flare. Their study also indicates that some of the evaporations observed in flare events may happen at a layer higher than the chromosphere or lower transition region. A probable mechanism for these evaporations is the rapid variation of downward heat flux from the hot reconnection outflow region.

\subsection{HXR and SXR views}

\begin{figure}[htbp]
\begin{center}
\includegraphics[width=\linewidth]{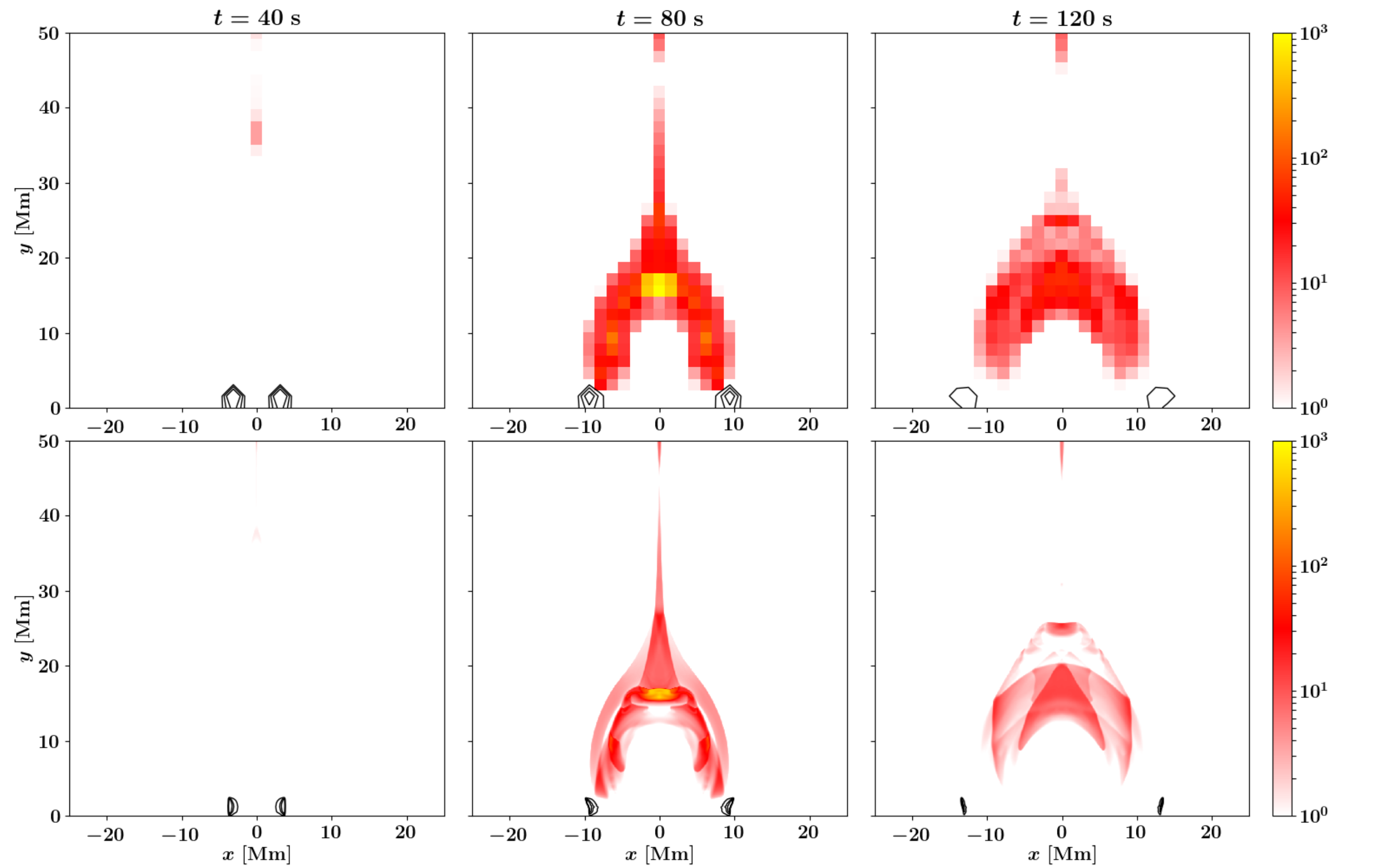}
\caption{Temporal evolution of SXR emission (6-12 keV, images) and HXR emission (25-50 keV, black contours) observed at 1 AU. The lower panels show the SXR and HXR emission at the simulation resolution, where the unit of SXR is photons arcsec$^{-2}$ cm$^{-1}$ s$^{-1}$ and the levels of the HXR contours are 0.1, 0.4 and 0.8 photons arcsec$^{-2}$ cm$^{-1}$ s$^{-1}$. The upper panels show the X-ray emission for a pixel size 2.15 arcsec $\times$ 2.15 arcsec, where the unit of SXR is photons pixel$^{-1}$ cm$^{-1}$ s$^{-1}$ and the levels of the HXR contours are 0.1, 0.5 and 1.0 photons pixel$^{-1}$ cm$^{-1}$ s$^{-1}$. The width in $z$-direction of the loop is assumed to be 20 Mm in the calculation of emission flux.}
\label{XR}
\end{center}
\end{figure}

\begin{figure}[htbp]
\begin{center}
\includegraphics[width=\linewidth]{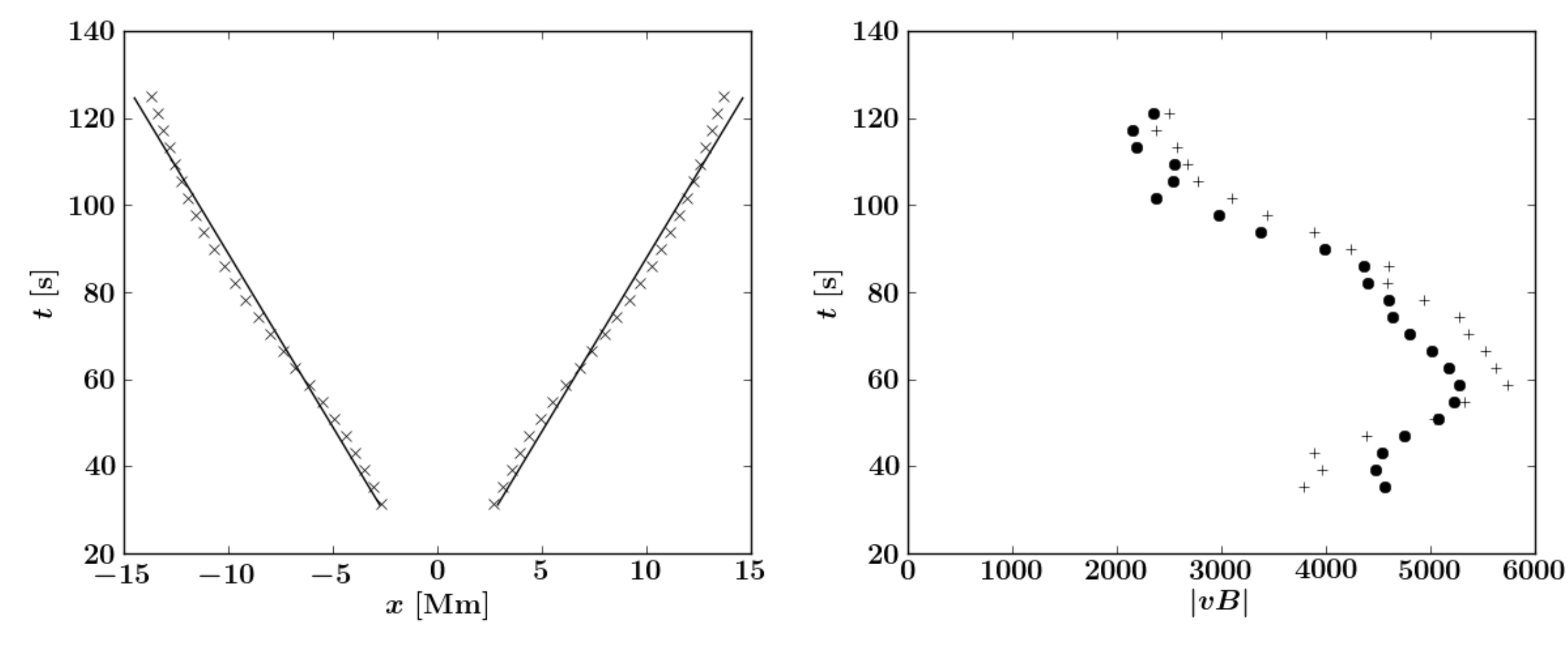}
\caption{Left: time-space plot of the footpoint HXR sources. The crosses show the $x$-coordinate of the HXR footpoint sources (in figure \ref{XR}). The straight lines are a linear fit of the tracks of the HXR sources and give a seperation speed of 125 km s$^{-1}$.  Right: dots show the values of $|v_{x} B_{y}|$ near the (relocating) reconnection point at different times and the crosses show the product of instantaneous footpoint HXR source speed and the chromospheric magnetic field magnitude $v_{foot} B_{foot}$.}
\label{vB}
\end{center}
\end{figure}

Here we analyze the X-ray emission of the simulated flare event. The thermal bremsstrahlung SXR emission is synthesized with the method reported in \citet{Pinto2015A&A} and the method to synthesize the non-thermal HXR emission is provided in Appendix~\ref{appHXR}. The SXR emission (6-12 keV) and HXR emission (25-50 keV) at $t=$ 40, 80 and 120 s are demonstrated in Figure~\ref{XR}. It clearly shows that a SXR loop and two HXR footpoint sources are formed. The features of the emissions fit those of the flare event reported in \citet{Masuda1994Natur} rather well: the SXR loop has a bright apex due to its high temperature and the HXR sources are located near the footpoints of the SXR loop. These HXR sources move in opposite direction during the simulation due to the change of the electron penetration site, in agreement with separating flare ribbon evolutions. The average seperation speed of our footpoint HXR sources obtained from the time-space plot is about 125 km s$^{-1}$, as shown in the left panel of figure~\ref{vB},. It has been suggested that the motion of flare ribbons and the photospheric/chromospheric flux swept out by them provides an observable measure of the reconnection rate, quantifying magnetic flux that gets convected into the diffusion region at the reconnection site (e.g. \citealp{Qiu2002ApJ}).  If this relationship holds, it provides a powerful means to quantify the coronal magnetic reconnection rate. Here we can, for the first time, confirm this relationship, by comparing the magnetic flux convection rate in the reconnection inflow region $v_{\perp,in} B_{\parallel,in}$, with the magnetic flux sweeping rate of the footpoint HXR source $v_{foot} B_{foot}$. This comparison is presented in the right panel of figure \ref{vB}. We found that the following equation is well satisfied throughout the entire flare process, relating the reconnection rate at the X-point with observable properties related to flare ribbons:
\begin{equation}
v_{\perp,in} B_{\parallel,in} = v_{foot} B_{foot} \,,
\end{equation}
where $v_{\perp,in}$ is the component of inflow velocity perpendicular to the reconnection current sheet, $B_{\parallel,in}$ is the component of the inflowing magnetic field that is parallel to the current sheet, while $v_{foot}$ and $B_{foot}$ are the motion speed and the magnetic field strength of the footpoint HXR sources. Note that $v_{foot}$ is in the $x$-direction and the magnetic field in the chromosphere $B_{foot}$ only has a $y$-component in our simulation. The ratio of $v_{foot} B_{foot}/v_{\perp,in} B_{\parallel,in}$ in the right panel of figure \ref{vB} ranges from $80\%$ to $130\%$. In principle, the motion of HXR footpoints is thus a more suitable instantaneous reconnection rate proxy than that provided by EUV flare ribbons, as fast electron deposition suffers much less delay than the effects provoked by thermal conduction.  For example, fast electrons arrive at the chromosphere 10 s earlier than the thermal conduction in figure \ref{field}, while it takes a time scale of 1 s for a fast electron with energy higher than 20 keV to move to the chromosphere from the coronal looptop located at a height of several tens of Mm. However, the observations of HXR emission are still restricted by a low temporal and spatial resolution. 

\begin{figure}[htbp]
\begin{center}
\includegraphics[width=\linewidth]{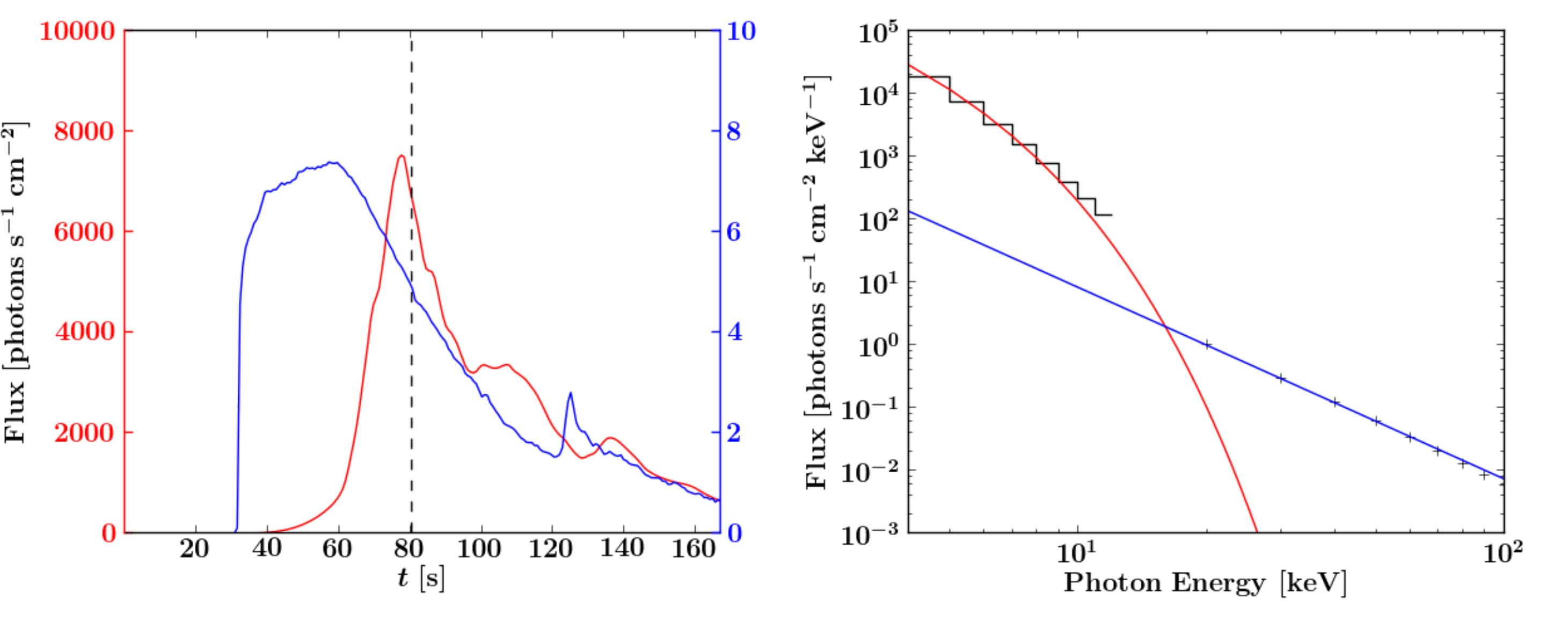}
\caption{Left: Temporal evolution of integrated HXR flux (blue) and integrated SXR flux (red). Right: X-ray spectrum at $t\approx 80\ \rm s$. The width in $z$-direction of the loop is assumed to be 20 Mm in the calculation of emission flux. The black line and the black crosses shows the SXR and HXR flux calculated from our simulation data. The red and blue solid lines show the results of fitting.}
\label{XRflux}
\end{center}
\end{figure}

The temporal profiles of the integrated SXR flux and integral HXR flux are displayed in Figure \ref{XRflux}, left panel. The figure shows the typical feature of the ``Neupert effect", namely that the SXR flux increases quickly when the HXR flux is high. This ``Neupert effect" is an empirical relationship between the HXR flux and SXR flux, stating that the temporal HXR flux has the same shape as the time derivative of the SXR flux \citep{Neupert1968ApJ, Hudson1991BAAS}. This effect is understood as follows: the energy deposition of fast electrons which produce the HXR flux is the main source to produce the hot plasma that is responsible for the thermal SXR emission. However, our Figure~\ref{field} clearly shows that the electron deposition fails to directly heat the chromospheric plasma to a high temperature. Our result supports the conclusion in \citet{Veronig2005ApJ}, where it is stated that fast electrons are not the main energy source of hot plasma supply and heating. The main source of the hot plasma in our simulation is actually from the reconnection outflow and the field aligned conduction. Therefore, we suggest that the nature behind some of the observed ``Neupert effect'' is that the non-thermal energy responsible for the HXR emission and the thermal energy responsible for the SXR emission have the same source, which is the reconnection process. This is suggested to work in those flare events where evaporations are not driven by fast electron deposition, or in events where the SXR emission is not determined by the fast-electron-driven evaporations, as in our simulation here. Those conditions are likely to appear in small flare events. For flare events where explosive evaporations are driven by non-thermal electrons, the energetic electron penetration is presumably still the main reason that leads to a ``Neupert effect''.

The spectrum of the X-ray (integral X-ray flux in the domain $-25\ \textrm{Mm} \leq x \leq 25\ \textrm{Mm}$ and $0\ \textrm{Mm} \leq y \leq 50\ \textrm{Mm}$) at t = 80 s is provided in the right panel of Figure \ref{XRflux}. The emission flux for photon energies lower than 12 keV is treated as thermal bremsstrahlung emission and is calculated with the plasma temperature and density, while the flux for photon energies higher than 20 keV is treated as non-thermal bremsstrahlung emission and is calculated from the evolving fast electron spectrum and plasma density. The integral SXR spectrum has a Gaussian distribution, while the HXR spectrum follows a power law shape. Fitting is performed to SXR and HXR spectrum with Equations 13.3.1 and 13.2.18 in \citet{Aschwanden2005psci}, respectively. The temperature obtained from fitting of the SXR spectrum is 17 MK, which is lower than the peak temperature of the flare loop (about 30 MK at the looptop). The corresponding emission measure is $8.9 \times 10^{47} \ \rm cm^{-3}$. The HXR spectrum has a spectral index of 3, which is lower than the spectral index $\delta=4$ of injected fast electrons at the apex. This result fits well the relationship between the fast electron spectrum and a chromospheric thick target bremsstrahlung HXR spectrum as predicted by \citet{Holman2011SSRv}.

\subsection{Looptop HXR sources and trapping}

The local HXR emission flux due to bremsstrahlung is in direct proportion to the local plasma density and the local fast electron flux. A looptop HXR source in observations can be comparable in magnitude to the footpoint sources, despite the fact that the footpoint plasma density is much higher than the looptop plasma density. Hence, similar source strengths likely appear when the fast electrons flux in the coronal looptop turns out to be much higher than in the chromospheric footpoints. Two reasons may lead to such a higher energetic electron flux: 1) fast electrons lose most of their energy in the looptop due to collisions and lead to a correspondingly lower flux in the high energy band at the footpoint; or 2) most fast electrons are trapped in the looptop and can not go into the footpoint.  In the case reported in the previous section, no looptop HXR source was found, consistent with the fact that particle trapping did not happen, turning the flare looptop into a thin target for untrapped fast electrons. Since our model also includes the possibility of particle trapping, we here demonstrate how a looptop HXR source can indeed be reproduced. To study the influence of fast electron trapping on the possible formation of a looptop HXR source, we vary the initial pitch angle for the fast electron population. 

\begin{figure}[htbp]
\begin{center}
\includegraphics[width=0.6\linewidth]{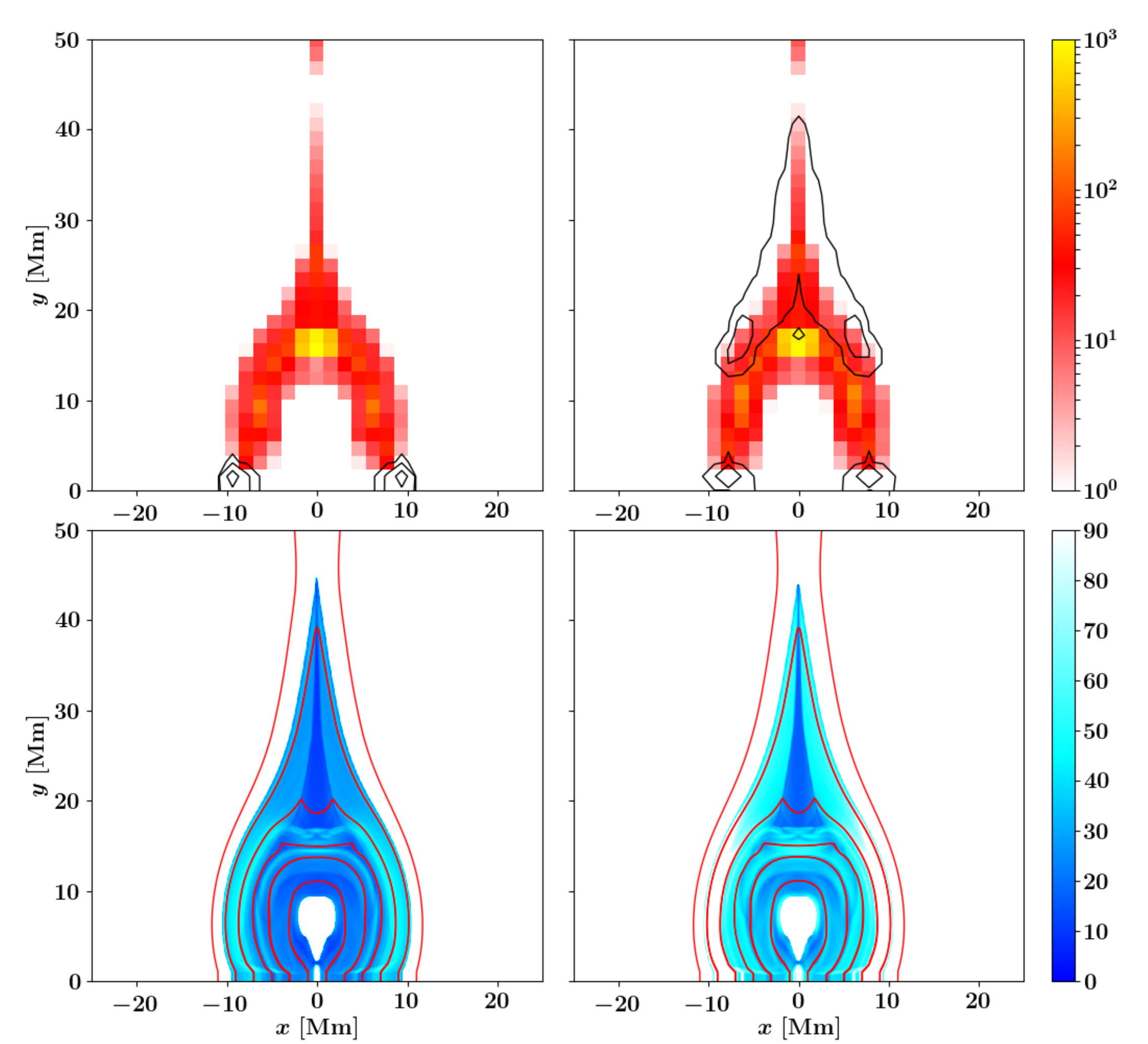}
\caption{HXR sources (black contours in the upper panels) at $t=80$ s for the case of initial pitch angle $18^{\circ}$ (left panels) and the case of initial pitch angle $30^{\circ}$ (right panels). Corresponding SXR emission (6-12 keV, photons pixel$^{-1}$ cm$^{-2}$ s$^{-1}$) are displayed in the upper panels and corresponding pitch angles (degree) are displayed in the lower panels. The levels of the contours are 0.01, 0.1 and 1.0 photons pixel$^{-1}$ cm$^{-2}$ s$^{-1}$. The size of one pixel is about 2.15 arcsec $\times$ 2.15 arcsec. The red lines in the lower panels are some selected magnetic field lines.}
\label{looptop}
\end{center}
\end{figure}

In this new case, the initial pitch angle of the fast electrons when they leave the acceleration site is set to $30^{\circ}$. In contrast, the initial pitch angle of the simulation reported in the previous section was set to $18^{\circ}$. With all other parameters unchanged, we find in this higher initial pitch angle scenario that the fast electrons do get trapped in the looptop due to the mirror force, since the magnetic field lines converge rapidly from the looptop to footpoints (see lower panels in Figure \ref{looptop}). As a result, a strong looptop HXR source is produced due to the local increase of the fast electron flux as shown in Figure \ref{looptop}. 
We also find that the footpoint HXR emission density in the new case is weaker than that in the first case, indicating that now the looptop is a thick target for trapped fast electrons.
The electrons are trapped in the looptop within a typical density of $N_e = 10^{10}\ \rm cm^{-3}$ for a timescale of $\tau_t = L_t / v_o = 15 \rm \ Mm/ 1500\ km/s = 10\ s$, where $L_t$ is the vertical length of the looptop (estimated from the contours of HXR emission in Figure \ref{looptop}) and  $v_o$ is the speed of the reconnection outflow, which is close to the Alfv\'en speed in the reconnection inflow regions. It means that the equivalent column depth of the looptop for an electron with a speed $v_{e}=10^{10}\ \rm cm\ s^{-1}$ is $ v_{e} N_e \tau_t \simeq 10^{21}\ \rm cm^{-2}$, which is enough for the electron to lose most of its energy. Hence, our model shows that the formation of a looptop HXR source can be a direct result of fast electron trapping.

A similar conclusion has recently been proposed by \citet{Kong2019ApJ}, as they suggest that the observed looptop HXR is a result of fast electron trapping. They suggest that the fast electrons are accelerated by the reconnection-driven termination shock and then are trapped in a small region above the top of the SXR loop and below the termination shock. In contrast to this, in our work we find that fast electrons, assumed to be accelerated in the reconnection current sheet, are trapped in a much larger region above and surrounding the termination shock. Both mechanisms may have a contribution to the formation of the observed looptop HXR sources, since both direct current field acceleration and shock acceleration are probable processes for the production of fast electrons in flare events. The mechanism proposed by \citet{Kong2019ApJ} is likely to produce a small but strong HXR source, while the mechanism suggested by us is likely to produce a larger, but weaker HXR source.

\section{Conclusion and discussion}\label{Concl}

In order to study important solar flare processes, like HXR source generation or details of chromospheric evaporation related to fast electron transport, we performed self-consistent multidimensional MHD simulations in which the fast electrons are included with an analytic model. Fast electrons are assumed to be accelerated in the reconnection outflow current sheet, move along magnetic field lines and lose their energy due to collisions. Thermal conduction, radiative losses, an initial VAL-C temperature profile, gravitational stratification and the non-local process of plasma heating due to the collision between fast electrons and the plasma are all incorporated in the model. The formation of a flare loop owing to reconnection is simulated ab initio and the generation of chromospheric plasma evaporation is reproduced. The thermal bremsstrahlung SXR emission and non-thermal HXR emission are synthesized based on plasma density, temperature and the instantaneous (and self-consistently evolving) fast electron spectra. We obtain the following main conclusions: 
\begin{itemize}
\item Chromospheric evaporations observed in solar flares can be triggered by thermal conduction rather than fast electron penetration, especially when the fast electron deposition rate is not extremely high;
\item The `Neupert effect', which describes the empirical relationship between the integral HXR and SXR flux, is qualitatively reproduced and derives physically from the fact that non-thermal and thermal energy evolutions are both linked to reconnection site processes;
\item The generation of a looptop HXR source is a result of fast electron trapping, as influenced by the initial pitch angle distribution of the accelerated particles;
\item Our simulation provides a conclusive demonstration that flare ribbon observations can indeed provide a good estimate of the varying reconnection rate at the X-point.
\end{itemize}

Note that high fast electron deposition rates, enough to trigger a strong evaporation, have been reported (e.g. \citealp{Milligan2006ApJ}). Here, we emphasize that our model with a generic flare temperature of $\sim$ 20 MK shows that thermal conduction on its own has the ability to produce strong evaporation of several hundreds km/s. This point has also been addressed by \citet{Polito2018ApJ} with a 1D hydrodynamic simulation performed by the RADYN code \citep{Carlsson1992ApJ,Carlsson1995ApJ,Carlsson1997ApJ}.

Our work has also some limitations, related to our handling of the fast electrons. Firstly, it is difficult to achieve a perfect energy conservation in the hybrid description, when coupling the MHD model to the analytic energetic electron model: we verified that the energy obtained from the acceleration site in the particle model equals about 70\% to 90\% of the energy removed from the MHD model. The deficit is intrinsic to the fact that our model energizes  particle beams as calculated from local plasma parameters interpolated to discrete points on a finite number of preselected (time-evolving) magnetic field lines, while the energy loss in the MHD model is calculated using a finite volume approach, where all quantities represent cell-average plasma parameters. Secondly, the treatment of the evolution of pitch angle is also simplified, where electrons of different energy are assumed to have the same pitch angle initially. This assumption can easily be relaxed by adopting an initial pitch angle distribution, but we here deliberately compared different cases with the same pitch angle, to show the trapping effect most clearly. Thirdly, the modeled chromosphere is assumed to be fully ionized in our simulation, which has smaller heat capacity than the partially ionized solar chromosphere. The simulated radiative loss is zero when the temperature is lower than $10^4$ K and it is treated as optically thin when the temperature is higher than $10^4$ K. However, the chromosphere should be optically thick for some spectral lines. To accurately evaluate chromospheric radiative loss, one needs to solve the radiative transfer and non-local thermal equilibrium, which is numerically challenging in multi-dimensional MHD simulations. A simple approach to get more realistic chromospheric radiative losses below a temperature of $1.5\times10^4$ K is proposed by \citet{Goodman2012ApJ}, in which the loss is a function of temperature and local plasma density. This method can be used in our future work. Fourthly, the identification of the acceleration site is based on our anomalous resistive MHD description, since this is difficult to directly obtain from a pure MHD model. A method to solve this problem would be to couple an MHD method with an embedded PIC region, with a treatment as done in \citet{Makwana2017CoPhC}, although the handling of heat fluxes may not be straightforward between the two regimes, and resorting to some analytic model as done here may still be required. 
Finally, the resolution of our 2.5D, grid-adaptive simulation (the smallest cells have a size of $\sim 48$ km) may not be high enough to resolve the transition region fully. The transition region has a small temperature length scale of order $\leq 1000$ m due to its large temperature gradient, which requires a cell size of $\leq 500$ m to resolve this region \citep{Bradshaw2013ApJ}. A poor resolution will lead to a changed balance between radiation and thermal conduction in the transition region, and then may lead to an underestimate of the coronal density and overestimate of coronal temperature. Since the numerical effort to achieve a high resolution simulation of cell size 500 m is quite high (and virtually impossible in multi-D large-scale setups), some methods have been developed to obtain a correct interaction between corona and chromosphere with a low spatial resolution, such as the TRAC method proposed by \citet{Johnston2019ApJ} and \citet{Johnston2020A&A}. This method broadens the temperature length scales in the unresolved transition region via increasing the local parallel thermal conductivity and decreasing the local radiative loss. One-dimensional numerical experiments in \citet{Johnston2019ApJ}  clearly demonstrate that accurate coronal densities, temperatures and velocities can be achieved by this method with a low resolution with grid cell of widths 125 km. However, this field-aligned method is still to be tested in multi-dimensional, grid-adaptive MHD simulations, where magnetic field lines would have to be traced from each pixel of the transition region throughout the simulation box and cutoff temperatures have to be evaluated for each field line. As magnetic field line tracing can already be achieved in our module, we plan to include this method into our model in the future.

In future work, the detailed treatment of chromospheric emission, the coupling to realistic flux emergence events, and the extension to full 3D radiative MHD including fast electron physics is also to be pursued. The fast electron treatment can be generalized to handle multiple beams with varying initial pitch angles, and can easily be adopted to full 3D scenarios. Further extensions of the model or the analysis of its results include the quantification of radio emission, and the inclusion of any (return) currents corresponding to the beams in the MHD feedback loop.

\section*{Acknowledgements}
RK was supported by a joint FWO-NSFC grant G0E9619N and received funding from the European Research Council (ERC) under the European Union’s Horizon 2020 research and innovation programme (grant agreement No. 833251 PROMINENT ERC-ADG 2018). WR received funding from the Chinese Scholarship Council.
This research is supported by Internal funds KU Leuven, project C14/19/089 TRACESpace. The
computational resources and services used in this work were provided by
the VSC (Flemish Supercomputer Center), funded by the Research Foundation
Flanders (FWO) and the Flemish Government - department EWI.

\appendix

\section{Plasma heating from fast electrons along a flux tube}\label{app1}

We will use a generalization of earlier 1D analytic models to quantify the heating rate along a time-evolving flux tube. Since the flare happens within seconds to minutes, we can reasonably assume that fieldline footpoints at low chromospheric to photospheric regions are fixed and line-tied, so that we can always identify the instantaneous field line as traced from those regions. At all times, our simulation box is then filled with a large collection of such flux tubes, whose axes are magnetic field lines traced from prechosen positions at our lower boundary. The local cross section of the entire flux tube element is calculated from flux conservation,  
\begin{equation}
A(s,t) B(s,t) = A_b B_b, \label{fc}
\end{equation}
where $s$ is the distance along the field line, $B(s,t)$ is the local magnetic field strength we obtain by interpolating from the MHD grid to the field line, $A_b$ and $B_b$ are the imposed cross-section and magnetic field strength at the lower boundary, and $A(s,t)$ is the local cross section we want to know along the field line. In our 2.5D simulation, only the magnetic strength in the $x-y$ plane is used and $A(s,t)$ denotes a width of the tube element in this $x-y$ plane, but the whole procedure carries over to 3D settings in a trivial way. On the basis of symmetry considerations, we will quantify arc lengths $s$ starting from a reference point $s_0$ where the field line crosses the $x=0$ vertical (aligned with gravity). Quantities at that reference point will be identified with a subscript zero, e.g. $B_0(t)$ and $A_0(t)$ for the instantaneous field strength and cross section, respectively. 

From the same reference position, which relocates as time progresses, we inject an instantaneous power law distribution of fast electrons, in practice given by an energy distribution according to
\begin{equation}
F_0 (E_0,t) = \frac{\mathcal{F}_0(t)}{E_c^2} (\delta-2) \left( \frac{E_0}{E_c} \right)^{-\delta} H(E_0 - E_c), 
\label{Fe0}
\end{equation}
where $H$ is the Heaviside step function, $E_c = 20 \ \rm keV = 3.2 \times 10^{-8} \ \rm erg$ is the fixed cutoff energy of the spectrum, and $\delta=4$ is the spectral index. Here $F_0$ quantifies the number of $\rm electrons\ cm^{-2}\ s^{-1}\ erg^{-1}$, while $\mathcal{F}_0(t)$ (in $\rm erg\ cm^{-2}\ s^{-1}$) quantifies the instantaneous energy flux injected into the flux tube element, obeying
\begin{equation}
\mathcal{F}_0(t) = \int_{0}^{\infty} E_0 F_0 (E_0,t) dE_0.
\end{equation}
The injected energy flux $\mathcal{F}_0(t)$ as well as the local instantaneous pitch angle $\mu_0(t)$ at the reference point are dynamically updated in a procedure outlined in Appendix~\ref{app3}, based on the energy budget evolution for the beam. In what follows, we assume that these two quantities are available for each field line at all times. 

The evolution of pitch angle of fast electrons in static non-homogeneous magnetic fields can be estimated from the conservation of magnetic moment, stating that ${v_{\perp}^2}/{B}$ remains constant,
which works between two collisions. During this period, the kinetic energy of the particle is conserved, leading to $(1-\mu^2)/B=const$, where $\mu=\cos \theta$ is the cosine of pitch angle $\theta$ between the beam velocity and the local field direction.
When we neglect the change of average pitch angle of the electrons in the collisions, this suffices to describe the evolution of the average pitch angle of the fast electron beam along each flux tube, since we can write
\begin{equation}
\mu(s,t) = \sqrt{1-\frac{B(s,t)}{B_0(t)}(1-\mu_0^2(t))}.
\label{mu2}
\end{equation}
A similar treatment of pitch angle is employed by \citet{McTiernan1990ApJ} in their analytic study of the behavior of non-thermal electron beams in astrophysical situations.

Following equation (23a) of \citet{Emslie1978ApJ}, the average energy loss rate of an electron beam traversing a fully ionized cold target plasma of electron number density $N_e$ (in $\rm cm^{-3}$) is given by
\begin{equation}
\frac{dE}{dt} = \frac{-2\pi e^4}{E} \Lambda_c N_e v, \label{emsl}
\end{equation}
where $v$  is the velocity (in $\rm cm\ s^{-1}$) of the incident electrons (initially $E=m_ev^2/2$), $E$ is the energy (in $\rm erg$) of the incident electrons, and $\Lambda_c$ denotes the Coulomb logarithm of the plasma. The value of $\Lambda_c$ is set to 25 in our simulation, a value appropriate for coronal conditions. Introducing the constant $K= 2\pi e^4$ and defining the column depth as 
\begin{equation}
N(s)=\int_{s_0}^{s} \frac{N_e(s')}{\mu(s')}  ds'  \,,
\end{equation}
we can rewrite Eq.~(\ref{emsl}) in terms of column depth, noting that $ds=v \mu dt$, thus $dN=N_e v\,dt$ to get a new expression of the average energy loss rate as function of column depth alone, namely
\begin{equation}
\frac{dE}{dN}=\frac{-K \Lambda_c}{E} \,.
\end{equation}
This is easily solved analytically, and we obtain 
\begin{equation}
E^2(E_0,N)=E_0^2-2K\Lambda_c N \,,
\label{EN}
\end{equation}
where $E_0$ is the energy at zero column depth. Note in particular that our definition of column depth differs from the one  in \citet{Emslie1978ApJ}, where it is merely using $N(s)=\int_{-\infty}^{s} N_e(s') ds'$ in his 1D analysis for a vertical beam down a uniform magnetic field.

In accord with \citet{Emslie1978ApJ}, the heating rate owing to collisions as a function of column depth $N$ can be given by
\begin{equation}
\tilde{H}_e(N,t)=\int_0^{\infty} F(E,N,t) \left| \frac{dE}{dN} \right| dE \,,
\label{heat}
\end{equation}
where $F(E,N,t)$ denotes the local electron flux distribution.
According to the conservation of electron flux throughout the flux tube, we get
\begin{equation}
F(E,N,t) A(N,t) dE = F_0(E_0,t) A_0(t) dE_0,
\label{FEN}
\end{equation}
where flux conservation means $A_0(t)/A(s,t)=B(s,t)/B_0(t)$ and we can easily revert from arc length $s$ to column depth $N(s)$. Considering this, the integration in equation~(\ref{heat}) becomes
\begin{equation}
\tilde{H}_e(N,t)=\int_0^{\infty} \left| \frac{dE}{dN} \right|(E_0,N) F_0(E_0,t) \frac{B(N,t)}{B_0(t)} dE_0 \,.
\end{equation}
Using Eqns.~(\ref{EN}) and (\ref{Fe0}), the result of the integration (in $\rm erg\ s^{-1}$) is given by
\begin{equation}
\tilde{H}_e(N,t) = \frac{1}{2} K \Lambda_c (\delta-2) \frac{B(N,t)}{B_0(t)} \frac{\mathcal{F}_0(t)}{E_c^2} \mathcal{B}_{x_c(N)} \left(\frac{\delta}{2},\frac{1}{2} \right) 
\left( \frac{N}{N_c} \right)^{-\frac{\delta}{2}},
\label{HeA}
\end{equation}
where $N_c = E_c^2/2K\Lambda_c$ is a constant, and $x_c(N)=2K\Lambda_c N/E_c^2$. In the integration, the incomplete beta function
\begin{equation}
\mathcal{B}_x (a,b) = \int_0^x y^{a-1} (1-y)^{b-1} dy
\end{equation}
is used.
Equation~(\ref{HeA}) has a similar form with the heating rate in \citet{Hawley1994ApJ}, with the difference coming from the treatment of pitch angle, as the change of pitch angle due to collisions has been included in their work. We neglect this change of pitch angle owing to collisions, but rather include the influence of magnetic field convergence and divergence, to incorporate trapping as explained in Appendix~\ref{app3}. The heating rate as a function of arc length along a field line is then given by 
\begin{equation}
H_e (s,t) = \tilde{H}_e(N,t) \frac{dN}{ds} = \tilde{H}_e(N(s),t) \frac{N_e (s,t)}{\mu(s,t)},
\label{Hes}
\end{equation}
where $H_e(s,t)$ has units of $\rm erg\ cm^{-3}\ s^{-1}$. This quantity needs to be remapped from the field lines to the MHD grid, to quantify $H_e(x,y,t)$ in our energy equation~(\ref{q-en}). This remapping is done as explained in Appendix~\ref{app4}.

\section{Magnetic field tracing}\label{app2}

The analytic fast electron heating model introduced in Appendix~\ref{app1} is in essence one-dimensional along a magnetic field line and the resulting heating rate is a function of distance along this field line. If we want to use it in a two-dimensional  (2D) or three-dimensional (3D) simulation, we need to trace magnetic field lines in 2D or 3D simulation boxes. To construct a 2D or 3D heating map $H_e(x,y,t)$ throughout the domain, we must trace enough magnetic field lines to sufficiently cover the block-adaptive hierarchical AMR grid used in the MHD simulation. Tracing magnetic field lines means knowing the coordinates of many points along each field line and obtaining the local plasma parameters at these points by linear interpolation from the AMR grid. Here we introduce our method to trace magnetic field lines in MPI-AMRVAC, where the generalization from the 2D case discussed to 3D is trivial. 

We provide the starting points of all field lines we trace and set the distance between two points on a field line to $\Delta l$, where $\Delta l$ equals the smallest size of any grid cell in the AMR simulation.
The starting points are distributed along the lower boundary of the simulation box and they distinguish the field lines. The tracing of a magnetic field line starts from these selected locations at the bottom boundary, and coordinates of successive points on the same field line are computed based on the coordinates and the magnetic field vector of the previous point. For example in 2D or a 2.5D setup, the coordinates of the $(i+1)$-th point on a field line are given by 
\begin{eqnarray}
x_{i+1} &=& x_i \pm \Delta l B_{x,i}/\sqrt{B_{x,i}^2+B_{y,i}^2}  \,, \\
y_{i+1} &=& y_i \pm \Delta l B_{y,i}/\sqrt{B_{x,i}^2+B_{y,i}^2}  \,,
\end{eqnarray}
where $\bm{B_i} = (B_{x,i},B_{y,i})$ is the (in-plane) magnetic field vector at the $i$th point. The sign $\pm$ determines whether the tracing direction is parallel or anti-parallel to the direction of the magnetic field. The value of $\bm{B_i}$ is obtained from linear interpolation of the magnetic field vector defined in the cell centers of the AMR grid, from a number of cells close to the point $(x_i,y_i)$. 

Since MPI-AMRVAC uses a dynamically evolving, block-adaptive grid, the simulation domain is divided into many blocks of different refinement levels at each time step in the MHD simulation. Each block contains a fixed number of cells of the same size, and the parallelism in the code implies that its information is only in memory for one of the typically many processors used to run a simulation. Of course, a magnetic field line will typically go through multiple blocks controlled by different processors. How to continue tracing when the field line goes from one block to another is a key technical implementation issue. In MPI-AMRVAC, every block has a block index ({\tt iblock}) and  processor index ({\tt ipe}) as used in the MPI communication. For each block, the code records the minimal and maximal boundary locations in each direction, as well as the {\tt ipe} and {\tt iblock} identifiers of all its neighbors. This information is used in the magnetic field tracing algorithm.   

\begin{figure*}[htbp]
\begin{center}
\includegraphics[width=0.8\linewidth]{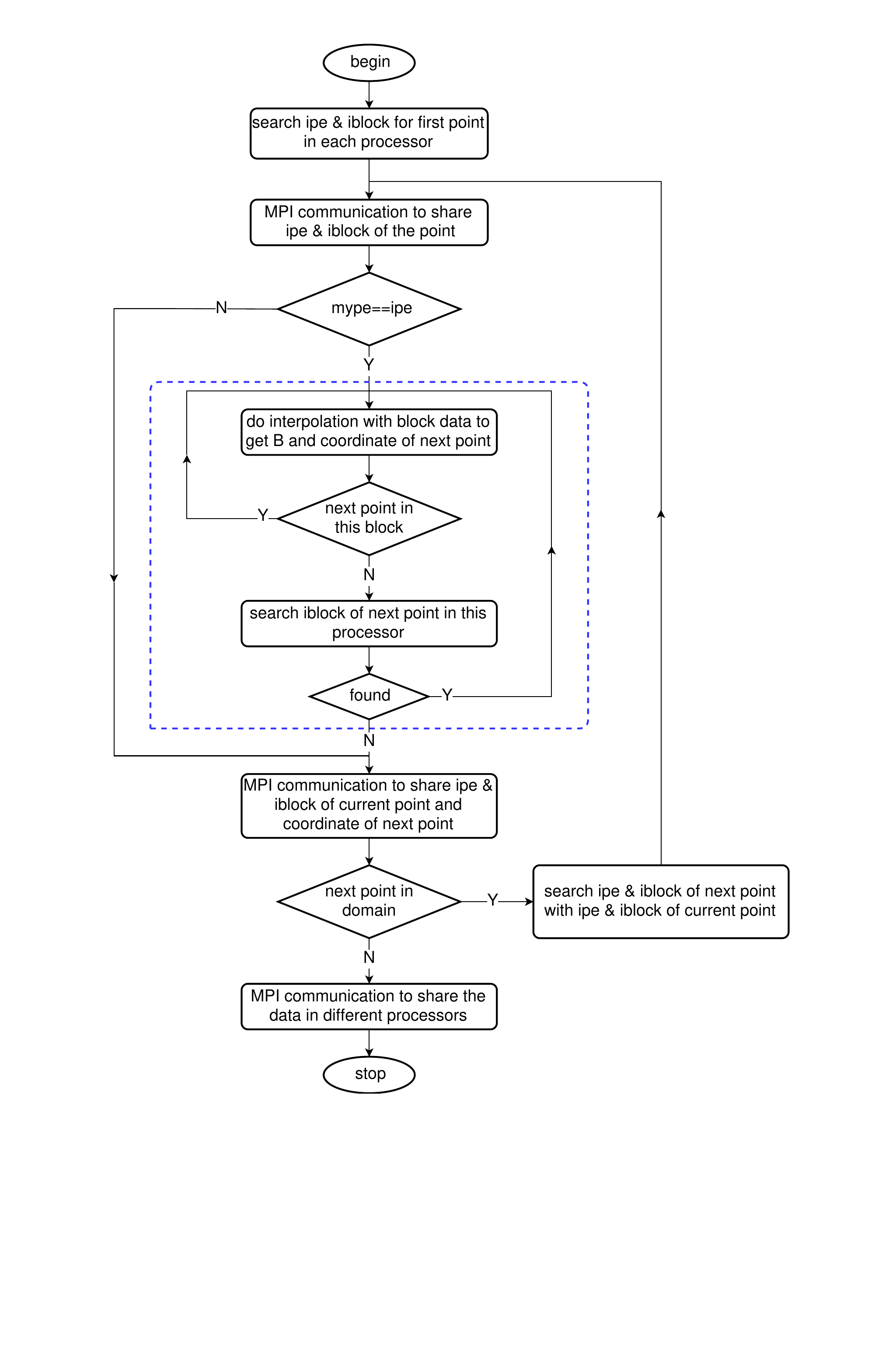}
\caption{Flow chart for magnetic field tracing in MPI-AMRVAC. The simulation domain is divided into multiple blocks controlled by different processors, where {\tt iblock} and {\tt ipe} indicate block and processor indices, respectively.}
\label{tracing}
\end{center}
\end{figure*}

Figure \ref{tracing} shows the flow chart to achieve magnetic field tracing in MPI-AMRVAC. We start with identifying in which block the starting point is located, based on the information of block boundaries: we will find a unique {\tt iblock} and corresponding processor {\tt ipe}. Thereafter, this info is shared to all other processors. As long as successive points in the same field line rely on data available to the local processor, we can continue tracing the same field line with local magnetic field vector and coordinate info, and this stops when the next needed point is not in any blocks controlled by the current processor. Then, the processor will communicate with all other processors to search the {\tt ipe} and {\tt iblock} of this next point from its known coordinates $(x_{i+1},y_{i+1})$.  Only one of the processors will have a corresponding {\tt ipe} and {\tt iblock} for it, and needs to inform all other processors. This processor will take over the work to further trace the field line. Tracing will stop when the field line goes outside of the simulation domain or a given number of points along the field line are already known. Communication among the processors must be performed to share the data of the points available in different processors, namely the coordinates of the points and the corresponding local plasma parameters (e.g. density) obtained from linear interpolation. 

We trace field lines serially with starting points at the lower boundary, distributed on both sides of the middle $(x,y)=(0,0)$. We stop tracing when the minimum distance between the field line and the line $x=0$ is larger than 5 Mm, since we are solely interested in magnetic field lines that go through the flare loop or reconnection region. While we build up the field lines, the coordinates $(x_i,y_i)$ and local plasma parameters like density and current density at these points are also obtained by interpolation. We recompute all magnetic field line data every $\Delta t_{flt} = \Delta l/v_r$, where $v_r= 5000\ \rm km\ s^{-1}$ is a flow value much higher than the reconnection outflow speeds we find typically. The distance between two neighboring starting points along the bottom boundary is also automatically adjusted in the simulation, to ensure that the maximum distance between two field lines is not very large, all along the field lines. The distance between two field lines is based on  flux conservation $A_b B_b = A(s) B(s)$ from equation~(\ref{fc}). We add a new starting point (and hence a new field line) between two previous starting points when the maximum distance as quantified by $\max{[A(s)]}$ is larger than $4 \Delta l$. Similarly, a starting point and field line is removed when the maximum distance is always smaller than $\Delta l$. 

\begin{figure*}[htbp]
\begin{center}
\includegraphics[width=0.6\linewidth]{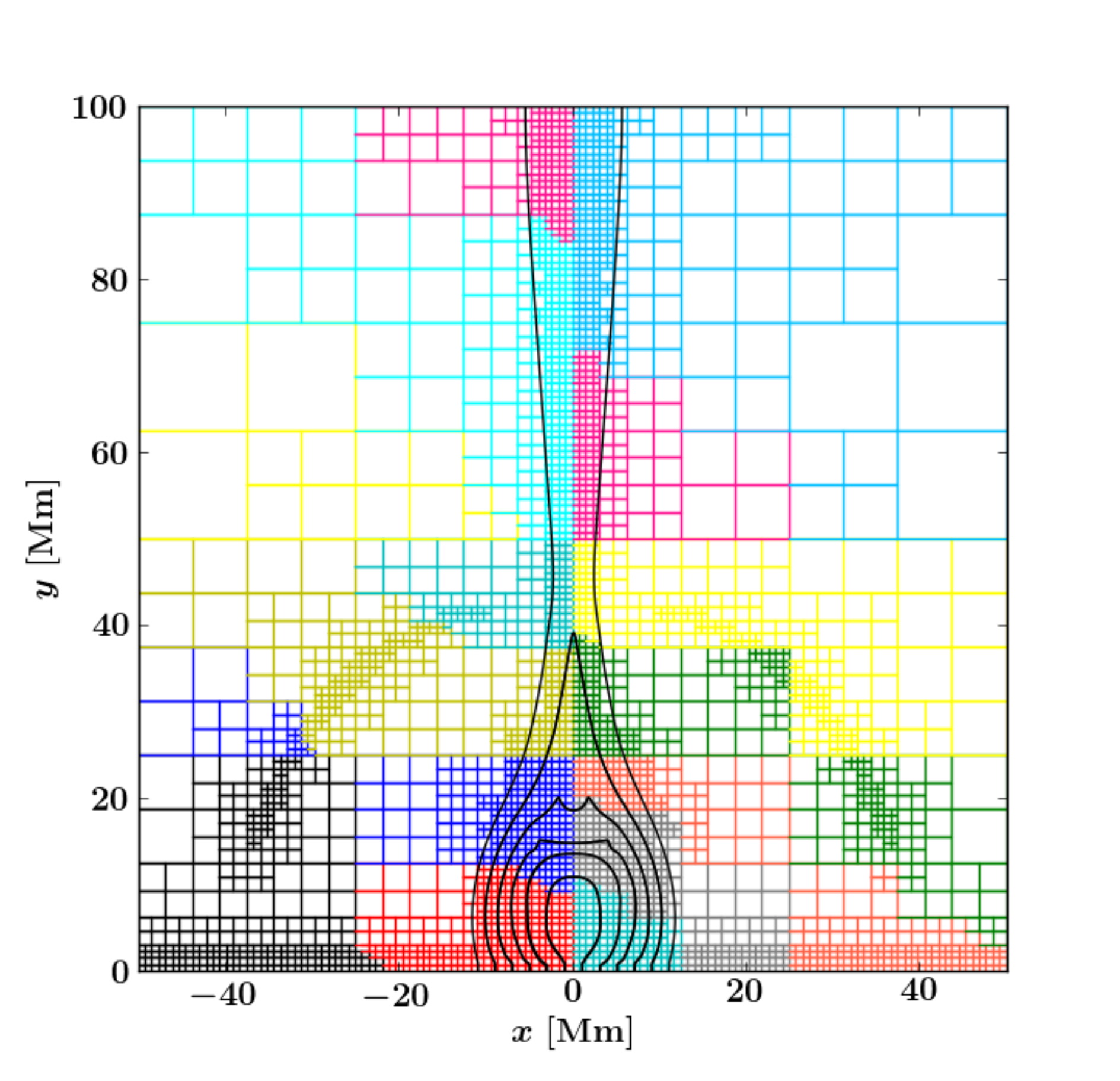}
\caption{Black lines demonstrate some of the magnetic field lines traced by us in the simulation. Squares of different colors are the boundaries of blocks controlled by different processors. Each block contains $16\times 16 = 256$ cells here.}
\label{block_field}
\end{center}
\end{figure*}

Usually, about 1000 field lines are traced during the simulation, where about half of them are traced parallel (anti-parallel) to the magnetic field with starting points at the left (right) haft of the lower boundary. For each field line, the locations of $\sim$ 2000 points along the field line are calculated and the corresponding density, velocity vector, pressure, magnetic field vector and current vector at these points are obtained from interpolation in the tracing. The results of magnetic field tracing in MPI-AMRVAC are displayed in figure \ref{block_field}.

\section{Dynamical evolution of initial pitch angle and energy flux}\label{app3}

The fast electrons present throughout a flux tube element fall into one of two categories: new electrons generated in this time step or remaining electrons that were generated in previous time steps. We need to account for both these populations, and use an energy budget argument to calculate the instantaneous energy flux $\mathcal{F}_0(t)$ at the reference point and the cosine pitch angle $\mu_0(t)$ at the reference point, and this for each field line.  Once known, the cosine pitch angle in other locations along the field line is given by equation~(\ref{mu2}). This equation determines if particle trapping happens on the field line, in between those locations where $\mu(s,t)$ decreases to zero before the fast electrons have left the simulation domain or lost their energy in the chromosphere. The symmetry implies that the possible vanishing of $\mu(s,t)$ leads to a symmetric $(s_{\rm left},s_{\rm right})$ pair in each field line which denote the reflection points. When particle trapping does not happen in the flux tube considered, $s_{\rm left}$ and $s_{\rm right}$ are located at the boundaries of the simulation box. For each field line, we then quantify a typical travel length $L_e(t)=s_{\rm right}(t)-s_{\rm left}(t)$. 
The average velocity of the fast electron population is a constant given by
\begin{equation}
v_{e,a} = \frac{\int_{E_c}^{\infty} F_0 (E_0,t) dE_0}{\int_{E_c}^{\infty} F_0 (E_0,t) /v(E_0) dE_0} 
= \frac{\frac{1}{2}-\delta}{1-\delta} \left( \frac{2E_c}{m_e} \right)^{\frac{1}{2}} \approx 10^5 \ \rm km\ s^{-1}, \label{vefix}
\end{equation}
where $v(E_0)=\sqrt{2E_0/m_e}$ as relativistic effects are neglected. Using $L_e(t)$ and $v_{e,a}$, a typical time scale is found, which we use to write $\mathcal{F}_0(t)$ as
\begin{equation}
\mathcal{F}_0(t) = \frac{v_{e,a} E_{tot}(t)}{A_0(t)L_e(t)},
\label{Fcalc}
\end{equation}
so that we now need to specify the evolution of the total energy carried by fast electrons in  our flux tube element $E_{tot}(t)$. This equation indicates that the energy flux across a cross section of a flux tube element $\mathcal{F}_0(t) \mu_0(t)$ equals the energy density $E_{tot}/(A_0 L_e)$ multiplied by the parallel velocity of the fast electrons given by $v_{e,a} \mu_0$. Note that the direction for the energy flux $\mathcal{F}_0(t)$ is thus not parallel to the magnetic field line when the pitch angle is not zero. For symmetry, we set the energy flux into both sides of the reference point $s_0$ (at $x=0$ on the field line) equal to $\mathcal{F}_0(t)$.

The evolution of the total energy carried by fast electrons in a flux tube element is governed by
\begin{equation}
\frac{d E_{tot}}{dt} = \left( \frac{d E_{tot}}{dt} \right)^+ - \left( \frac{d E_{tot}}{dt} \right)^- \,, \label{bal}
\end{equation}
where
\begin{equation}
\left( \frac{d E_{tot}}{dt} \right)^+ = \int Q_e (s,t) A(s,t) d s 
\end{equation}
represents the total energy for newly generated electrons in the flux tube, where $Q_e(s,t)$ is given by equation~(\ref{joule}), but evaluated along the field line. Note that this corresponds to the energy reduced from the MHD system.
In the balance~(\ref{bal}), the $\left( \frac{d E_{tot}}{dt} \right)^-$ is the energy loss caused by Coulomb collisions, which must relate to integration of the heating caused due to collisions along the tube element. Accounting for the possibility of trapping, this energy loss rate should be set to
\begin{equation}
\left( \frac{d E_{tot}}{dt} \right)^- = \int_{s_0}^{s_{\rm left}} H_e (|s-s_0|) A(s) d|s| + \int_{s_0}^{s_{\rm right}} H_e (|s-s_0|) A(s) d|s|,
\end{equation}
where $H_e(s,t)$ is the heating rate as a function of arc length as given by equation~(\ref{Hes}). In practice, this introduces a complication since the energy flux $\mathcal{F}_0(t) $ and the cosine pitch angle $\mu_0(t)$ at the reference point are required for the calculation of the heating $H_e(s,t)$.  Therefore, we discretize equation~(\ref{bal}) into a time marching scheme given by
\begin{equation}
E_{tot} (t) = \Delta E_{tot}^+ (t) + E_{tot} (t-\Delta t) - \Delta E_{tot}^- (t-\Delta t) \,, \label{bald}
\end{equation} 
where 
\begin{equation}
\Delta E_{tot}^{\pm} (t) = \left( \frac{d E_{tot}}{dt} \right)^{\pm} \Delta t \,.
\end{equation}
Note that the energy loss due to collisions in a previous time step is used in~(\ref{bald}), thereby resolving the complication we mentioned. The time marching starts from a zero value for $E_{tot}$ (and $\Delta E_{tot}^-$).

To complete the calculations, we still need to calculate the cosine pitch angle variation $\mu_0(t)$.
This makes the following assumption: newly accelerated fast electrons should have an average perpendicular kinetic energy of $m_e v_{\perp,a}^2/2$ and an average energy of $m_e v_{e,a}^2/2$ when leaving the acceleration site, where both $v_{\perp,a}^2$ and $v_{e,a}^2$ are constants, with the latter given by equation~(\ref{vefix}) and the former fixed by prescribing a fixed input average beam pitch angle. Note that the acceleration site is an extended region, namely everywhere where $Q_e (s,t)$ is not zero, surrounding the reference point at $s_0$.
Again relying on an argument based on the first adiabatic invariant, we can compute $v_{\perp,0,new}^2$ for newly accelerated electrons in a time step at the reference point, by averaging $v_{\perp}^2 / B$ with weight $Q_e (s,t)$ and then multiply by the magnetic field strength at the reference point $B_0(t)$, i.e.
\begin{equation}
v_{\perp,0,new}^2 = B_0(t) v_{\perp,a}^2 \frac{ \int \frac{Q_e(s,t) A(s,t)}{B(s,t)}  ds}{\int Q_e(s,t) A(s,t) ds} \,.
\end{equation}
Thereafter, we average the squared perpendicular velocity at the reference point of these newly accelerated fast electrons with the remaining fast electrons already accelerated in the previous time steps, as weighted by energy, as in
\begin{equation}
v_{\perp,0}^2 (t) = \frac{B_0 (t)}{E_{tot} (t)} \left \{ \frac{v_{\perp,0,new}^2}{B_0(t)} \Delta E_{tot}^+ (t) + \left[ E_{tot} (t-\Delta t) - \Delta E_{tot}^- (t-\Delta t) \right] \frac{v_{\perp,0}^2 (t-\Delta t)}{B_0 (t-\Delta t)}  \right \} , 
\label{pupdate}
\end{equation}
which requires us to record the historic values of $E_{tot}-\Delta E_{tot}^-$ and $v_{\perp,0}^2/B_0$ of each field line from the previous time step. In an analogous fashion, the squared parallel speed for the newly accelerated electrons comes from $v_{\parallel,0,new}^2=v_{e,a}^2-v_{\perp,0,new}^2$, and is used to update the total squared parallel speed accounting for new and already present fast electrons as 
\begin{equation}
v_{\parallel,0}^2 (t) = \frac{1}{E_{tot} (t)} \left \{ v_{\parallel,0,new}^2 \Delta E_{tot}^+(t) + \left[ E_{tot} (t-\Delta t) - \Delta E_{tot}^- (t-\Delta t) \right] v_{\parallel,0}^2 (t-\Delta t)  \right \} \,
\end{equation}
such that we also need to record the previous value of $v_{\parallel,0}^2$. Thereafter, the cosine pitch angle can be easily obtained with
\begin{equation}
\mu_0 (t) = \sqrt{ \frac{v_{\parallel,0}^2 (t)}{v_{\parallel,0}^2 (t)+v_{\perp,0}^2(t)}}.
\end{equation}
Note that when there is no new energy injection, i.e. when $\Delta E_{tot}^+ (t)=0$, equation~(\ref{pupdate}) ensures $v_{\perp,0}^2/B_0 = const$. The procedure outlined here allows the perpendicular velocity of fast electrons coming from the acceleration site to change due to mirror forces when they move along magnetic field lines.

\section{Interpolation from magnetic field lines to the MHD AMR grid}\label{app4}

The local heating rate along given points on the traced magnetic field lines can be obtained as discussed in Appendix~\ref{app1}. To obtain the reaction of the plasma to the heating in the MHD simulation, we need to interpolate this heating rate from field lines to the grid cells, quantifying the source term $H_e (x,y,t)$ in the governing equations. There are several technical difficulties in this interpolation. One derives from the data structure of the AMR-MHD simulation: the simulation domain is divided into many blocks that contain multiple cells, with different spatial cell sizes in different blocks, and this block information is spread across different processors in the MPI environment. The other difficulty comes from the fact that the coordinates of the points on the many available field lines where we know the local heating rate are not regularly spaced. As a result, we can not use simple interpolation methods like bi-linear interpolation easily. 

\begin{figure*}[htbp]
\begin{center}
\includegraphics[width=\linewidth]{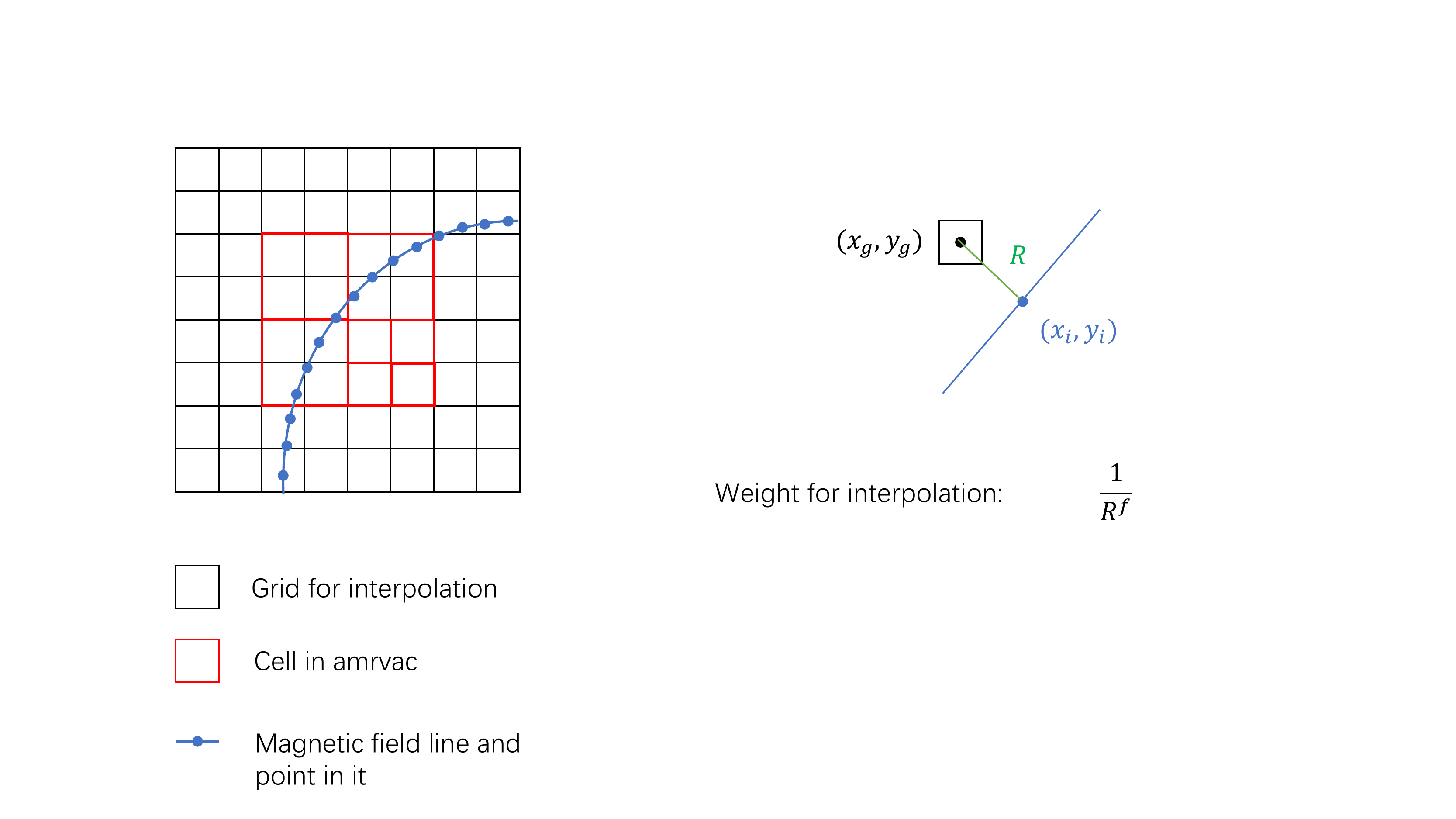}
\caption{The main algorithmic ingredients employed for interpolation between field lines and AMR grid. Left: the use of a uniform background grid table. Right: the anti-distance interpolation.}
\label{interpolation}
\end{center}
\end{figure*}

To solve the first difficulty, before the interpolation is performed, we create a fixed high resolution uniform background grid table in which the cell size equals the minimal available MHD cell size, such that the MHD cell boundaries match the background grid boundaries (figure \ref{interpolation}). This uniform background grid allows a more efficient interpolation from the field lines to its uniform cells. Once this is done,  the heating rate for each AMR-MHD grid cell can be obtained by a more direct summation. The size of the domain covered by the background uniform grid table can be smaller than the simulation box, but the regions where heating is important should be covered by the table. The region where $-40\ \textrm{Mm} \leq x \leq 40\ \textrm{Mm}$ and $0\ \textrm{Mm} \leq y \leq 60\ \textrm{Mm}$ is covered by the table in our simulation.

\begin{figure*}[htbp]
\begin{center}
\includegraphics[width=\linewidth]{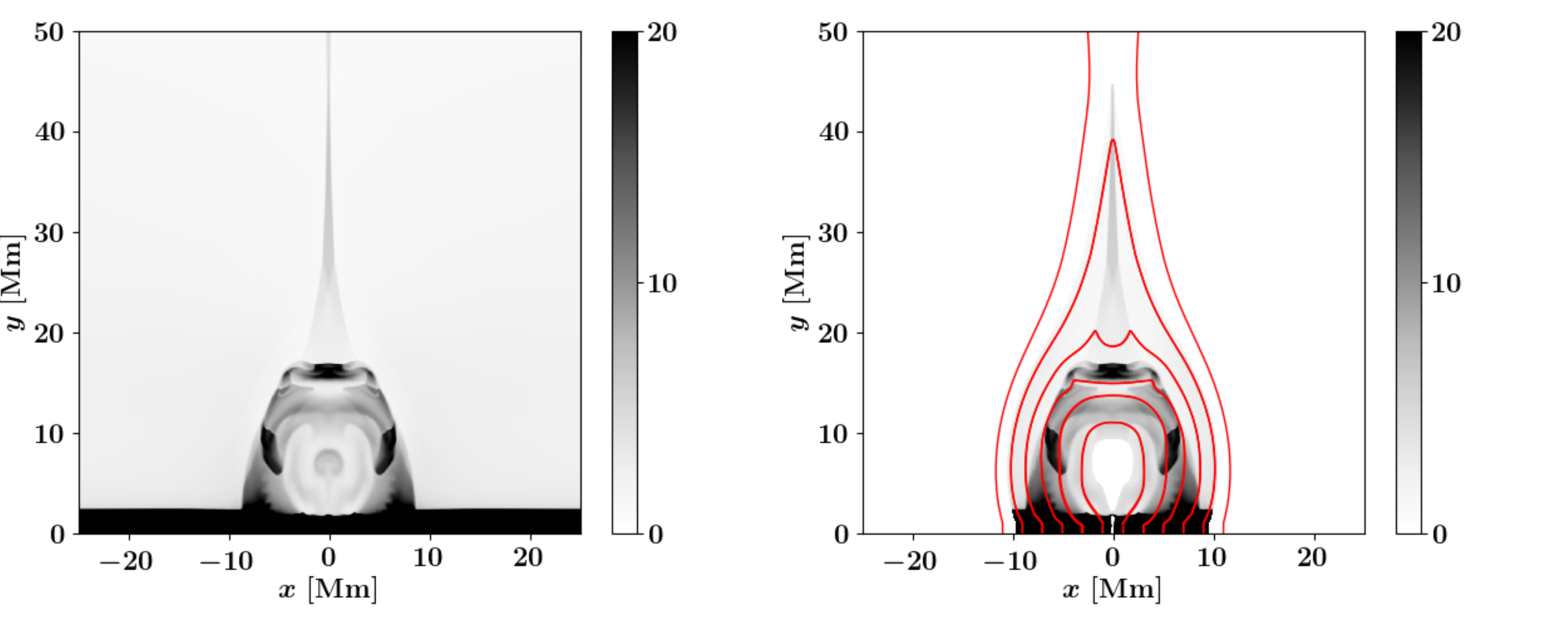}
\caption{Left: original density. Right: interpolated density. Red lines show part of the magnetic field lines that we used for the interpolation. The interpolation in our simulation only focus on the reconnection downflow and the flare loop where there may be fast electrons. For the field lines that do not go through those regions, the values of the interpolated variable in the field lines are set to zero.}
\label{interpresults}
\end{center}
\end{figure*}

To solve the second difficulty, anti-distance interpolation is employed as illustrated in figure \ref{interpolation}. The weight of interpolation is $w_i=1/R_i^f$, where $R_i$ is the distance between a background table grid center and a point on the magnetic field line, where $f=4$ is employed in our simulation. We traverse the magnetic field lines, and for each point $(x_i,y_i)$ on field lines, we find all table grid cell centers $(x_g,y_g)$ where $|x_i-x_g|<dx$ and $|y_i-y_g|<dy$. Thereafter, we record the cumulative weights for each grid cell $(x_g,y_g)$, i.e. for the interpolation $w=\sum_i 1/R_i^f$ and the contribution of the heating rate $H=\sum_i w_i H_i$, where $H_i$ is the local heating at the point and $w_i=1/R_i^f$. After traversing all the field lines, we can calculate the background grid heating rate from $H (x_g,y_g)=H/w$. The parameters $dx$ and $dy$ are not constant, and are taken larger in the regions where the magnetic field is weak, since the distance between two field lines can become large in such regions. In our simulation, we set $dx=dy=\Delta l B_b/B(x_p,y_p)$, where $\Delta l$ is the distance between two points in a field line, $B(x_p,y_p)$ is the local magnetic field strength (in-plane) and $B_b=35\ \rm G$ is the strength at the lower boundary. Where $B_b/B(x_p,y_p)>3$, we set $dx=dy=3 \Delta l$. Where $R_i<0.01\Delta l$, we set $w_i=1/(0.01\Delta l)^f$. The interpolation only needs to traverse the points in magnetic field lines one time. The interpolation results are displayed in figure \ref{interpresults} and demonstrate that small structures are well reproduced during the interpolation from field lines to grid.

\section{Hard X-ray synthesis}\label{appHXR}

The HXR emission in solar flares is determined by non-thermal bremsstrahlung. We synthesize HXR emission based on the treatment of fast electrons in the heating model introduced in appendices~\ref{app1} and \ref{app3}. The local HXR spectrum is given in emissivity by
 \begin{equation}
j(\epsilon)=\int_{\epsilon}^{\infty} \frac{N_i {F_l(E)}}{\mu} \mathcal{Q} (\epsilon,E) dE,
\label{je}
\end{equation}
where $N_i$ is the local proton number density, $\mu$ is cosine pitch angle and $F_l$ (in $\rm electrons\ cm^{-2}\ s^{-1}\ keV^{-1}$, identical to $F$ appearing in equation~(\ref{heat}) except for the energy unit used) is the local fast electron spectrum. Furthermore, $\mathcal{Q}$ is the Bremsstrahlung cross-section, $\epsilon$ is photon energy and $E$ denotes the electron energy. The Bethe-Heitler Bremsstrahlung cross-section is adopted \citep{Kontar2011SSRv}, which gives
\begin{equation}
\mathcal{Q} (\epsilon,E)=Z^2 \frac{\sigma_0}{\epsilon E} \ln \frac{1+\sqrt{1-\epsilon/E}}{1-\sqrt{1-\epsilon/E}},
\end{equation}
where $\sigma_0=7.9 \times 10^{-25} \ \rm cm^{-2}\ keV$ and $Z$ is ion charge.  Note that the Bethe-Heitler Bremsstrahlung cross-section has a simple expression and performs well in non-relativistic conditions. The relativistic effects can not be neglected when the incident electrons have a high energy, of order 100 keV. A better approximation can be found in \citet{Haug1997A&A} for such relativistic conditions. Haug's approximation has a much more involved expression, but is very useful in the numerical computation of HXR emission.

\begin{figure*}[htbp]
\begin{center}
\includegraphics[width=0.8\linewidth]{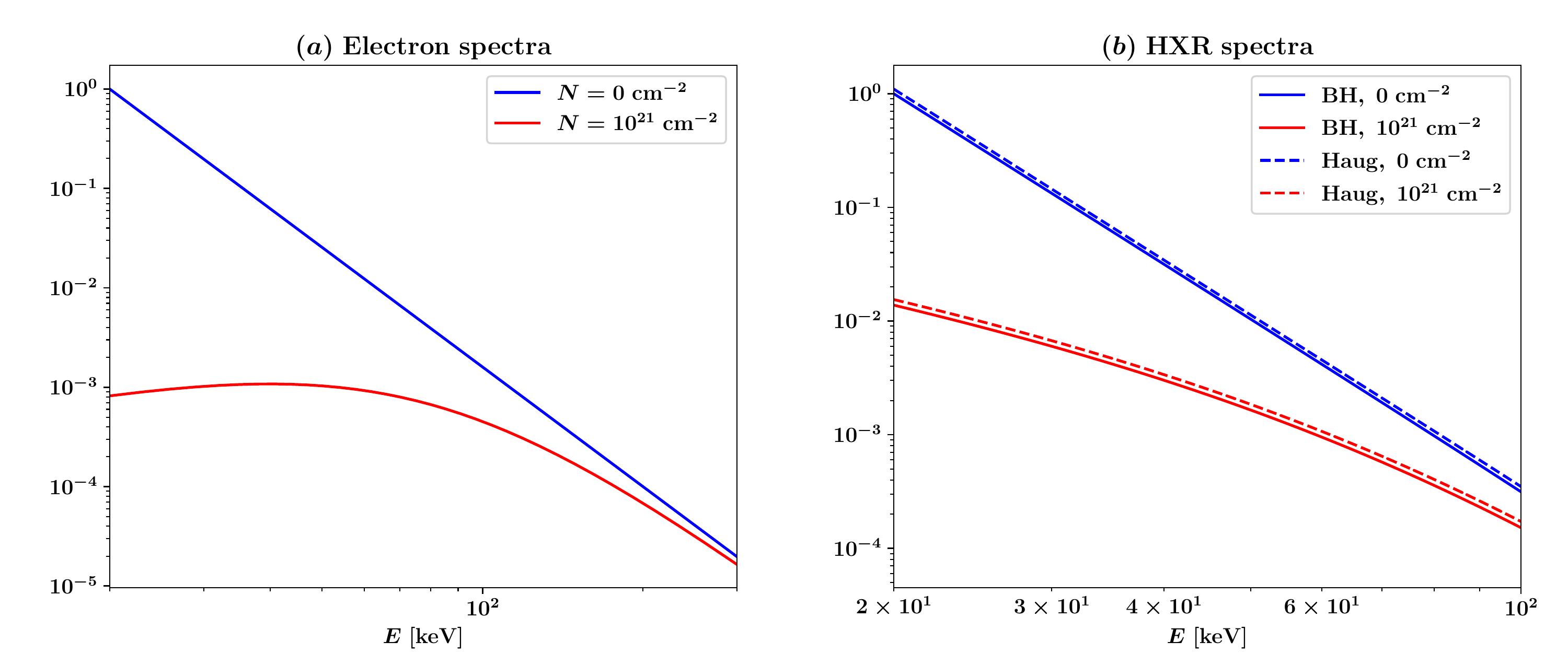}
\caption{Fast electron (left) and HXR spectra (right) for different column depths. The solid lines in the right panel are the HXR fluxes calculated with the Bethe-Heitler Bremsstrahlung cross-section (Equation (29) in this paper), while the dashed lines are the fluxes calculated with the combination of Equation (4) and (5) in \citet{Haug1997A&A}. }
\label{spectra}
\end{center}
\end{figure*}

The treatment of fast electrons in the HXR synthesis model uses the fast electron deposition model we explained earlier. Its initial energy distribution follows a power law distribution given by equation \ref{Fe0} and the distribution is assumed to change when the electrons move along magnetic field lines and collide with background plasma. The evolution of the distribution is governed by equations (\ref{Fe0}), (\ref{EN}), (\ref{mu2}) and (\ref{FEN}). The local energy distribution in an energy range $\rm (20\ keV, 300\ keV)$ is calculated by dividing the energy range into 280 energy bins. The electron flux $F_l=\sum_b F_b$ in each bin $\rm (E_b, E_b +1\ keV)$ is given by 
\begin{equation}
F_b = \frac{B}{B_0} \int_{E_l}^{E_u} F_0 (E_0) dE_0 = \frac{1-\mu^2}{1-\mu_0^2} \int_{E_l}^{E_u} F_0 (E_0) dE_0,
\end{equation}
where $E_l=\sqrt{E_b^2 +2K\Lambda_c N}$ and $E_u=\sqrt{(E_b+1)^2+2K\Lambda_cN}$. Figure \ref{spectra} shows the fast electron spectra for $20-300\ \rm keV$ at column depths $N=0$ (at injection) and $N=10^{21}\ \rm cm^{-2}$ and the corresponding HXR spectra between $20-100\ \rm keV$.

\end{document}